\documentclass[preprint,journal]{vgtc}            

\onlineid{1298}

\ieeedoi{10.1109/TVCG.2023.3326931}

\vgtccategory{Research}

\title{Visual Analysis of Displacement Processes in Porous Media using Spatio-Temporal Flow Graphs}

\author{%
  \authororcid{Alexander Straub}{0000-0002-6749-9710},
  \authororcid{Nikolaos Karadimitriou}{0000-0002-9461-6214},
  \authororcid{Guido Reina}{0000-0003-4127-1897},
  \authororcid{Steffen Frey}{0000-0002-1872-6905},
  \authororcid{Holger Steeb}{0000-0001-7602-4920}, and 
  \authororcid{Thomas Ertl}{0000-0003-4019-2505}%
}

\authorfooter{
  \item
  	Alexander Straub, Nikolaos Karadimitriou, Guido Reina, Holger Steeb, Thomas Ertl are with University of Stuttgart, Germany\\
  	E-mail: alexander.straub@visus.uni-stuttgart.de
  \item
  	Steffen Frey is with University of Groningen, The Netherlands.
}

\abstract{%
  We developed a new approach comprised of different visualizations for the comparative spatio-temporal analysis of displacement processes in porous media.
  We aim to analyze and compare ensemble datasets from experiments to gain insight into the influence of different parameters on fluid flow.
  To capture the displacement of a defending fluid by an invading fluid, we first condense an input image series to a single time map.
  From this map, we generate a spatio-temporal flow graph covering the whole process.
  \revision{This graph is further simplified to only reflect topological changes in the movement of the invading fluid.}
  Our interactive tools allow the visual analysis of these processes by visualizing the graph structure and the context of the experimental setup, as well as by providing charts for multiple metrics.
  We apply our approach to analyze and compare ensemble datasets jointly with domain experts, where we vary either fluid properties or the solid structure of the porous medium.
  We finally report the generated insights from the domain experts and discuss our contribution's advantages, generality, and limitations.
}

\keywords{Comparative visualization, ensemble, graph, porous media.}

\teaser{%
  \centering%
  \subfloat[\label{fig:time-triangular}]{%
    \includegraphics[width=0.33\columnwidth,trim={0cm 0.7cm 0cm 0cm},clip]{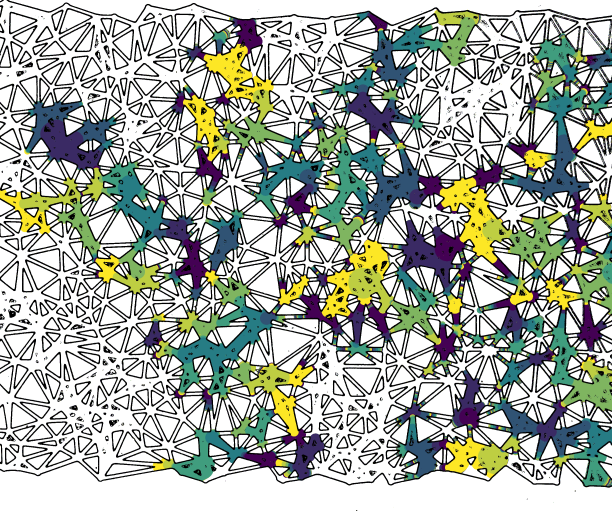}%
  }%
  \hfill%
  \subfloat[\label{fig:graph-triangular-simplified}]{%
    \includegraphics[width=0.33\columnwidth,trim={0cm 6.4cm 0cm 6.4cm},clip]{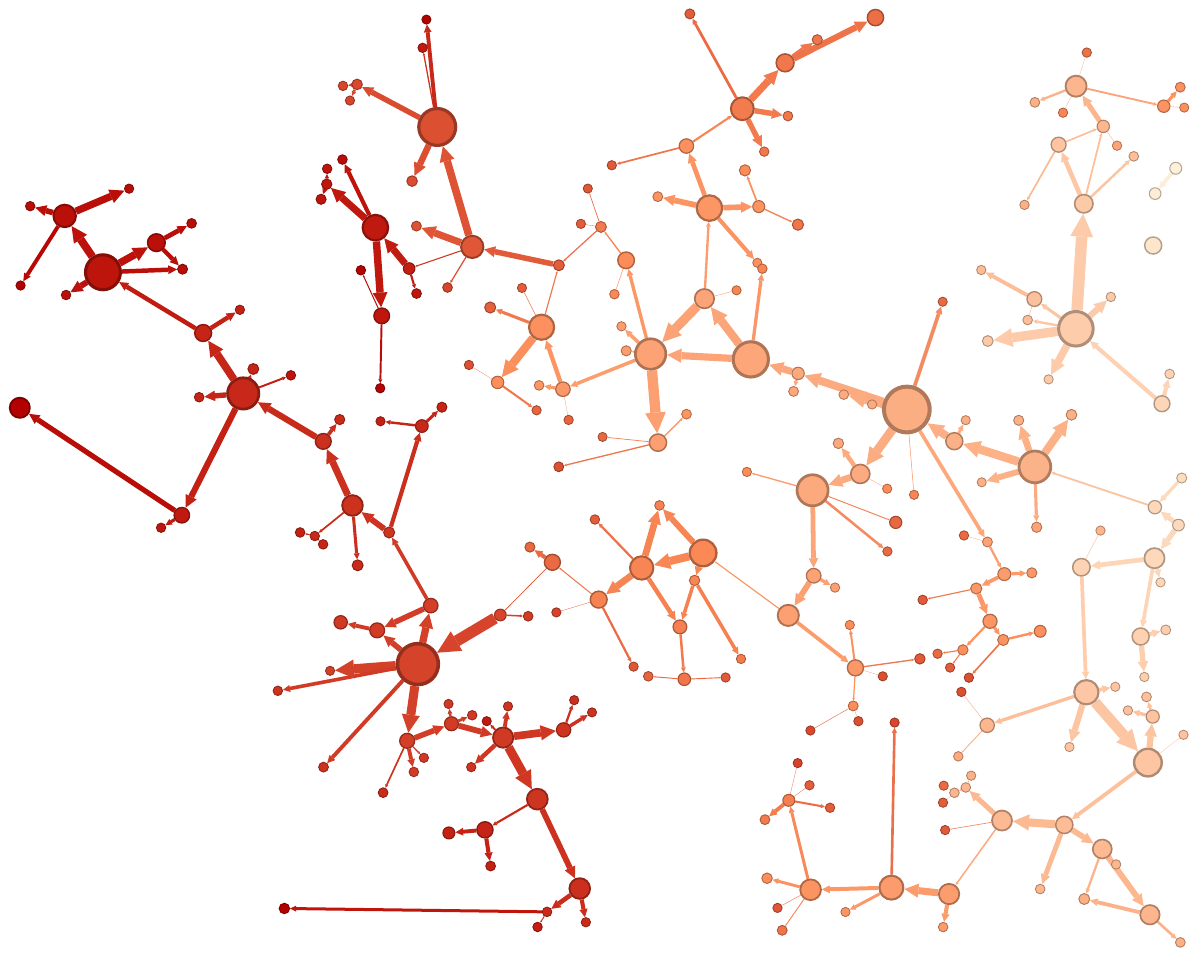}%
  }%
  \hfill%
  \subfloat[\label{fig:graph-triangular-mainchannel}]{%
    \includegraphics[width=0.33\columnwidth,trim={0cm 6.4cm 0cm 6.4cm},clip]{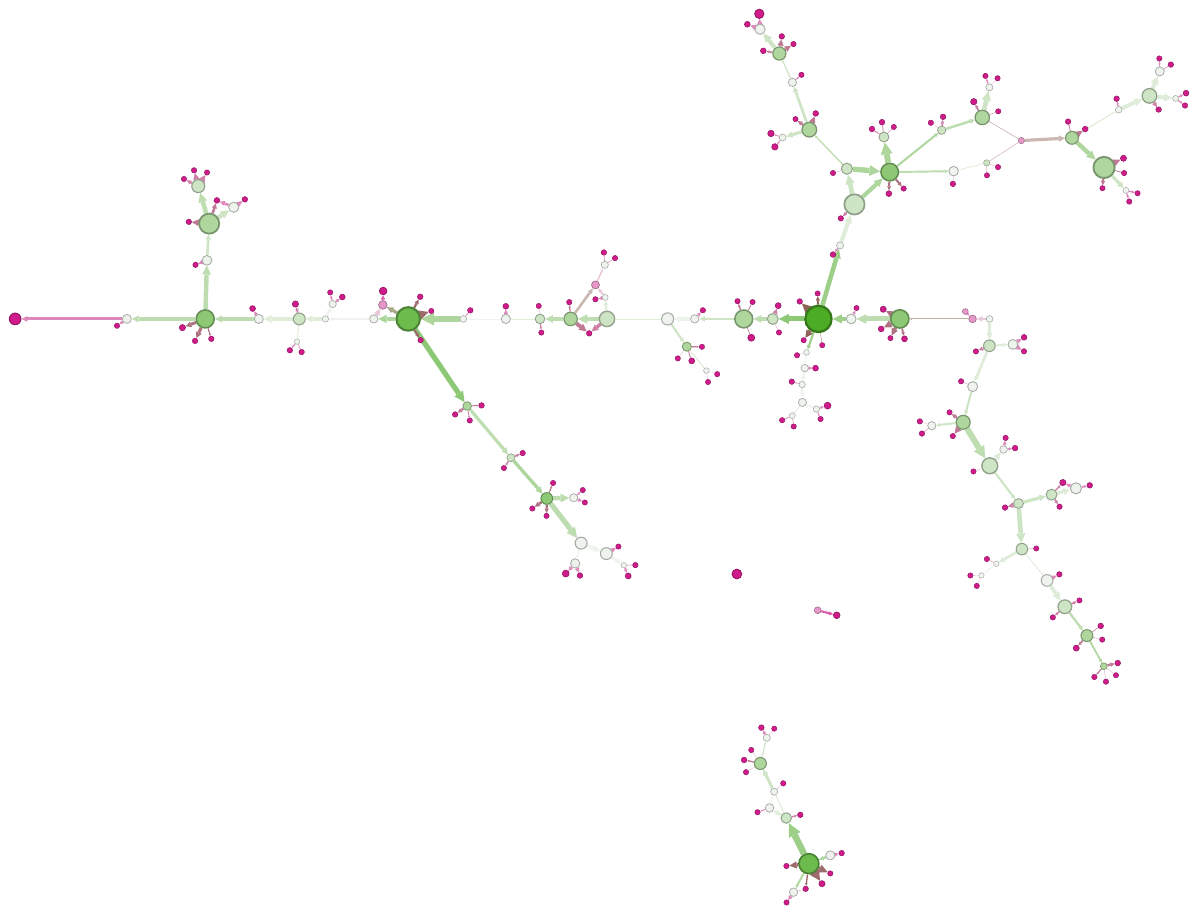}%
  }%
  \caption{%
    Abstraction levels of fluid flow through a porous medium with a triangular solid structure.
    \subref{fig:time-triangular}~Time map showing the progression of the invading fluid within the medium, with void space~(white), solid structures~(black outlines), and time information for the invading fluid.
    Time is mapped to a uniform periodic color map~(\viridis) where each period covers the same physical time duration.
    Large velocities can be observed when filling large pores, and small velocities occur when moving through throats.
    \subref{fig:graph-triangular-simplified}~Displacement graph as abstraction, with node positions at the barycenters of flow fronts, time mapped to color~(\reds), area mapped to node size, and velocity mapped to edge width.
    \subref{fig:graph-triangular-mainchannel}~Breakthrough graph with abstracted layout showing the main channel from inlet to outlet on a horizontal line.
    Only relative distances between nodes on the main channel are identical to~\subref{fig:graph-triangular-simplified}.
    Branches, as well as disconnected parts of the graph, are easily distinguishable.
    Color is now mapped to the degree of outgoing edges~(\categories) to further highlight branching.
  }%
  \label{fig:teaser}%
}

\renewcommand{\manuscriptnotetxt}{\vspace{4em}}

\graphicspath{{figs/}{figures/}{pictures/}{images/}{./}} 

\usepackage{mathptmx}                  

\usepackage[svgnames]{xcolor}
\usepackage{datetime}
\usepackage{amsmath}
\usepackage{amsfonts}
\usepackage{tabu}
\usepackage{booktabs}

\definecolor{gblue}{RGB}{66,133,244}
\definecolor{gyellow}{RGB}{244,160,0}
\definecolor{gred}{RGB}{219,68,55}
\definecolor{ggreen}{RGB}{15,157,88}

\newcommand{\revision}[1]{#1}

\definecolor{plotblue}{RGB}{97,159,202}%
\definecolor{plotorange}{RGB}{255,165,85}%
\definecolor{plotgreen}{RGB}{106,188,106}%

\newcommand{\viridis}{\smash{\resizebox{3em}{1ex}{\includegraphics[width=30em,height=10em,trim=1 30 1 2,clip]{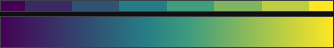}}}}
\newcommand{\reds}{\smash{\resizebox{3em}{1ex}{\includegraphics{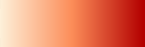}}}}
\newcommand{\categories}{\smash{\resizebox{3em}{1ex}{\includegraphics{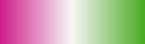}}}}

\renewcommand{\vec}[1]{\mathbf{#1}}

\newcommand{\MatrixX}[1]{\begin{matrix}#1\end{matrix}}

\usepackage{enumitem}

\usepackage{pifont}

\newcommand{\One}{172}
\newcommand{\Two}{173}
\newcommand{\Three}{174}
\newcommand{\Four}{175}

\newcommand{\circledNumber}[2]{\scalebox{#1}{\ding{#2}}}

% ...
\usepackage{amssymb}
\usepackage{siunitx}
\DeclareDocumentCommand{\tbf}{m}{#1}
\DeclareDocumentCommand{\illcomp}{mm}{\textbf{\textcolor{#1}{#2}}}
\DeclareDocumentCommand{\illcomp}{mm}{\textit{#2}}

\DeclareDocumentCommand{\ca}{}{\text{Ca}}

\DeclareDocumentCommand{\m}{}{\text{M}}

\begin{document}

\newcommand{\ensembleParams}[2]{\text{Ca}=10^{#1},~\text{M}=#2}

\newcommand{\segmThres}{\beta}

\firstsection{Introduction}

\maketitle

Two-phase flows and especially displacement processes in porous media---materials with interconnected void spaces---are of high importance for various industrial and environmental applications like enhanced oil recovery~\cite{Alzahid2017}, soil remediation~\cite{Khan2004}, and inkjet printing~\cite{Kettle2010}. 
However, the underlying physical processes have not fully been understood, yet, and significant experimental work is conducted in porous media research to shed light on this.

In this work, we discuss our visual analysis approach for the comparative spatiotemporal analysis of the forced displacement of a fluid saturating the void space with another fluid injected into the porous medium in a laboratory setting.
Depending on the parameters of the conducted experiments and the porous structures, the behavior of the liquids varies significantly.
One of the metrics used to characterize flow is the capillary number.
However, even though the capillary number is perceived as a metric for the boundary conditions of flow, there has been no consensus related to its effectiveness, both in terms of definition and use.
Most of the criticism relies on the fact that even though a global flux may be applied during a primary drainage event, the local capillary number, meaning the pore scale velocity, can vary depending on the local geometrical constraints.
In order to test this hypothesis, we proceed with the visual analysis of several primary drainage events for various boundary and physical conditions and pore space geometries.
A set of 11 experiments for different boundary and physical conditions is revisited from a previous publication~\cite{Frey2021}, as well as two nearly identical pore networks with a minimal variation between them, and a triangle-based network are employed under capillarity-dominated primary drainage conditions.
From the collected images, we anticipate extracting information on the pore-scale effects during primary drainage and how the physical properties assigned to the process, like pore-scale velocity and interfacial area, correspond to the globally assumed values.
The effect of the grain geometry on pore scale observations and their potential mismatch with average properties was investigated. 

We propose a novel process-driven approach to generate a spatio-temporal graph via temporal tracking of fluid advancement and introduce techniques to use it for visual analysis.
This is the joint work of visualization experts and scientists from porous media research.
The graph generation substantially improves upon the work by Frey et al.~\cite{Frey2021} as it applies to arbitrary geometry, captures additional information, and is generally more robust.
This is crucial for dealing with potentially noisy and misaligned experimental data.
We show the analyses both approaches can perform, the limitations of the novel approach, and the additional analyses and insights that have been enabled.

%

\section{Related Work}

This work is most closely related to the approach presented by Frey et al.~\cite{Frey2021}.
As the base of their approach, they extract a pore network graph, which allows to classify the whole image space into solids, pores, and throats.
This data structure is employed to extract all metrics used in the analysis.
One drawback of this approach is that the solid needs to be convex for it to work.
We never explicitly extract the solid and do not distinguish between pores and throats.
This makes our approach robust and generally applicable also to arbitrary geometries.
Additionally, we visualize velocity information and produce a graph that is more fine-grained and follows the flow, as opposed to the porous structure.
Other similarities and differences are discussed in more detail in~\autoref{sec:discussion}.

\textbf{Visualization of Porous Media and Liquid Phases.}
Previous work in visualization on porous media primarily focuses on the rendering of porous media.
Examples are different shader-based approaches~\cite{Grottel2010} or rendering in VR and demonstrating how to deal with occlusion~\cite{Naumov2014}.
An analysis of pore geometries in the context of CO\textsubscript{2} sequestration is contributed by Zhang et al.~\cite{Zhang2019}.
This is based on X-ray CT scans of liquid-filled sandstone, adding automatic classification of CO\textsubscript{2} bubbles and correlating these bubbles regarding morphological and geometric similarities.
The investigation of microfluidics from confocal microscopy recordings~\cite{Winter2021} is related to our work.
However, this only considers single-phase flow and rectilinear pore space structures.
The particle simulation ensemble investigated in the 2016 SciVis Contest~\cite{Geveci2016} shares viscous fingering with the experiments we investigate in this paper.
However, this only regarded interaction between liquids.
A comprehensive system for hierarchically exploring the ensemble was shown by Gralka et al.~\cite{Gralka2018}, starting from overall metrics and drilling down into ensemble members to investigate finger topology and vortex structures.
An approach based on topological analysis was contributed by Favelier et al.~\cite{Favelier2016}.
Here, fingers are extracted to derive and visualize metrics for comparison.
An extension by Lukasczyk et al.~\cite{Lukasczyk2017} integrates tracking graphs~\cite{Widanagamaachchi2012} and a query interface coupled with a database for ensemble analysis.
Glyphs that represent high-level parameters of ensemble members were presented by Luciani et al.~\cite{Luciani2019} to complement 3D visualizations.
Soler et al.\cite{Soler2019} combine geometry and topology:
they present a novel metric based on persistence diagrams to compute the agreement of viscous fingering simulations with a ground-truth dataset.
Reeb graphs have also been employed to extract finger structures and their skeletons~\cite{Xu2022}.
Here, glyphs give an abstracted overview of viscous and gravitational fingers.
Additionally, the visual analysis system uses a tracking graph and allows to interactively explore both spatial and abstract representation.
\revision{Similarly, Reeb graphs have been applied to porous media to find persistent paths represented in flow graphs~\cite{Ushizima2012}.
Other work frequently focuses on extracting the pore network graph, which represents the connected void space of the porous medium and classifies it into pores and throats.
While Frey et al.~\cite{Frey2021} provide a simple algorithm for 2D pore network extraction, Aboulhassan et al.~\cite{Aboulhassan2015} extract a \emph{backbone} graph from three-dimensional solar cell materials.
We refer to Heinzl and Stappen~\cite{Heinzl2017} for a survey on material science, which includes visualization of porous media.}

\textbf{Ensemble Visualization.}
Ensemble data are recognized to be generally challenging to analyze~\cite{Obermaier2014}, and several techniques have been proposed to tackle specific applications.
\revision{Several surveys give an overview of the state of the art regarding multifaceted data~\cite{Kehrer2013}, ensemble data~\cite{Wang2019}, and parameter space analysis~\cite{Sedlmair2014}.
Early examples include approaches for the analysis of climate simulation ensembles regarding statistics~\cite{Potter2009} and uncertainty~\cite{Sanyal2010}, and a system for ensemble steering~\cite{Waser2010}.}
Metrics for comparison and exploration of the simulation space include squared differences~\cite{Bruckner2010}, regional joint variance~\cite{Hummel2013}, and the tracking of coherent regions via optical flow~\cite{Kumpf2019}.
Shape similarity for particle data has been contributed by Hao et al.~\cite{Hao2016} and He et al.~\cite{He2020} compute surface distances employing surface density estimates.
Employing graphs as an abstraction for individual datasets of an ensemble, they have to be somehow compared.
For this, metrics~\cite{Wills2020} or graph spectra~\cite{Wilson2008} can be used, or they can be compared visually~\cite{Andrews2009}.

\textbf{Porous Media Research.}
In the literature, many models are proposed and applied for two-phase flow.
One of the most commonly applied models is proposed by van Genuchten~\cite{Genuchten1980}, where capillary pressure is a non-linear function of the wetting phase saturation.
Extended theories for two-phase flow include the inclusion of specific interfacial areas between the fluid phases as a separate state variable~\cite{Hassanizadeh1993}, the inclusion of process variables that depend on higher-order morphological features~\cite{Kurzeja2014}, and the assumption that capillary pressure crucially also depends on its spatial distribution in the porous domain~\cite{Schlueter2016}, which can be expressed with the first three Minkowski functionals \textit{M0--M2}~\cite{Armstrong2019}.
From an experimental perspective, two-phase flow has been studied both with natural and artificial porous media.
\revision{For purposes of realism in the experiments, samples extracted from real rocks have been used to analyze $\text{CO}_2$ storage~\cite{Andrew2013}, to investigate the effect of low salinity~\cite{Bartels2016}, and to observe Haines jumps~\cite{Berg2013}.}
\revision{However, for reasons of simplicity and better parameter control, techniques have been developed to manufacture artificial porous media with well-defined physical and surface properties.
They were used to lay the foundations for Darcy's law~\cite{Darcy1857}, as well as Hele-Shaw flow~\cite{Shaw1898}.
Later works employing such artificial media focused on investigating the steady state and its equations~\cite{Grosser1988} and on solute transport~\cite{Corapcioglu1997}.}
In the case of manufactured microfluidic chips used in two-phase flow studies, the term micromodel was introduced~\cite{Karadimitriou2012} to describe artificial, transparent porous media with "on-demand" average properties---like pore size distribution and porosity---, pore-scale features on the scale of microns, and overall dimensions on the centimeter scale.
The general categories for the fabrication method of micromodels are etching processes~(either chemical or ion)~\cite{Mattax1961} or lithography~\cite{Thompson1983}.
Soft-lithography is widely employed due to its comparably low cost and high precision~\cite{Xia1998}.
The microfluidic chip, i.e., the micromodel, used in this work was produced with soft lithography out of Poly-Di-Methyl-Siloxane~(PDMS).
Lenormand et al.~\cite{Lenormand1988} proceeded with the categorization of two-phase flow, and more specifically, primary drainage, in the capillary number~$\ca{}$--viscosity ratio~$\m{}$ space.
Based on this, they categorize the flow regimes into \emph{capillary fingering} for low~$\ca{}$, \emph{viscous fingering} for increased $\ca{}$ and low $\m{}$, and \emph{stable front} for elevated $\ca{}$ and $\m{}$.
As the means to investigate the validity of the proposed extended theories for two-phase flow, Cheng et al.~\cite{Cheng2004} performed sequential drainage and imbibition cycles experimentally to investigate the role of the interfacial area between fluids as a separate state variable under quasi-static flow conditions.
Karadimitriou et al.~\cite{Karadimitriou2014} extended their research with experiments with micromodels under dynamic conditions~(using a PDMS micromodel).

\section{Fundamentals of Porous Media} 
\label{sec:fundamentals}

\begin{figure}[t]%
    \centering
  \includegraphics[width=0.9\linewidth,trim={0 0 0 0.8cm},clip]{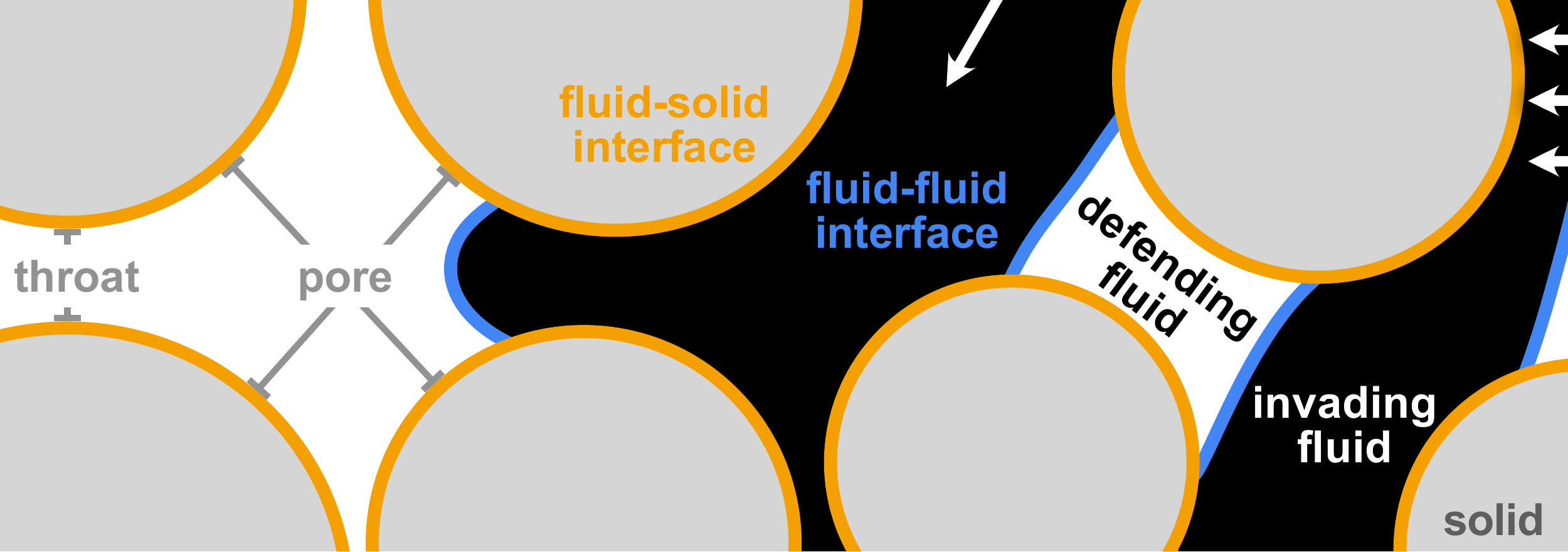}%
  \vspace{6pt}%
  \caption{%
    Illustration of porous medium terminology (adapted from \cite{Frey2021}).
  }%
  \label{fig:basics}%
\end{figure}

A porous medium contains solid material as well as a distribution of void space.
\tbf{Porosity} depicts the void volume ratio against the porous medium's bulk volume.
The void space is classified into \illcomp{pore}{pores}~(large void space) and \illcomp{throat}{throats}~(narrow spaces connecting the pore bodies).
In a porous medium, pore bodies and pore throats are generally interconnected, allowing flow through the porous medium.
Further terminology for porous media is depicted in~\autoref{fig:basics}.

In this work, we study the forced displacement of a \illcomp{wetting}{wetting fluid} by an incoming \illcomp{nonwetting}{non-wetting fluid} in a porous medium.
Here, the concept of wettability and the corresponding differentiation of the fluids into \tbf{wetting} and \tbf{non-wetting} phases describes the affinity of a fluid towards the solid surface, which stems from inter-molecular interactions at the \illcomp{fluidsolid}{fluid-solid interface}, where the \tbf{non-wetting fluid} has a lower affinity---i.e., weaker adhesive forces---towards the solid than the \tbf{wetting fluid}.
The displacement process is referred to as drainage~(with the term imbibition denoting the opposite).
In an experimental setup, a key event commonly referred to as a breakthrough is the formation of an uninterrupted connection of displacing fluid between its entry into the system on one side and its exit on another.

Two kinds of forces are the most relevant for displacement processes in a porous medium.
\emph{Capillary forces} cause slow fluid displacement dictated exclusively by the {geometry} of the pore space (\tbf{pores} and \tbf{throats}).
With imbibition, capillarity potentially drives flow in narrow spaces even against gravity and other forces (e.g., as can be observed with paper absorbing water).
This \illcomp{capflow}{capillary flow} is due to tension at the \illcomp{fluidfluid}{fluid-fluid interface} and the wetting properties regarding the solid phase at the \tbf{fluid-solid interface}.
\emph{Viscous forces} are induced by friction between the fluid and solid walls, slowing down the flow at the boundary of a fluid and creating layers of different velocities.
These cause shear stresses parallel to these layers, propagating the slowing effect from solid walls to inner fluid layers.
This results in \illcomp{lamflow}{laminar flow}, i.e., fluid flow in internal layers of different velocities.
Internal friction between the layers of a fluid and the impact on its flow is expressed via viscosity.
This is proportional to the velocity gradient between layers.
For high velocities, the flow is called \illcomp{viscdriven}{viscosity-driven}.
%
%
Two numbers are commonly employed to characterize the flow.
The \emph{capillary number}~$\ca{}$ is a dimensionless quantity representing the relative effect of \tbf{viscous forces} against \tbf{capillary forces}.
It is commonly expressed as the product of viscosity and fluid velocity divided by interfacial tension.
Roughly speaking, for low capillary numbers~($\lesssim 10^{-5}$), flow in porous media is dominated by \tbf{capillary forces}, whereas for high capillary numbers~($\gtrsim 10^{-3}$), the \tbf{capillary forces} are negligible compared to \tbf{viscous forces}.
Intermediate capillary numbers result in a combination of both forces.
The \emph{viscosity ratio}~$\m{}$ describes the relationship between the viscosities of the invading fluid versus the defending fluid.

\revision{For information about the experiment setup, we would like to refer the interested reader to Section 3 of the paper by Frey et al.~\cite{Frey2021}.
Due to the intrinsic pseudo-2D nature of such artificial porous media, with the third dimension being reduced to a few micrometers, we can treat the experiment as 2D without significant information loss.
However, there has been criticism regarding the use of artificial 2D porous media and their representativeness in comparison to real (or better, natural) materials.
The immediate effect of the lack of the third dimension on the so-called REV scale is on the total permeability of the sample.
However, the physics of the processes involved do not change.
Hence, such approaches are used from a fundamental science perspective: 2D representations are feasible to start with and easier to interpret.}

\section{Method}
\label{sec:method}

\begin{figure}[t]%
  \centering%
  \def\labelHeight{0.35cm}%
  \begin{minipage}[b][\labelHeight][t]{0.04\columnwidth}%
    \subfloat[\label{fig:datasets-circular}]{ }%
  \end{minipage}%
  \begin{minipage}{0.96\columnwidth}%
    \includegraphics[width=0.16\columnwidth]{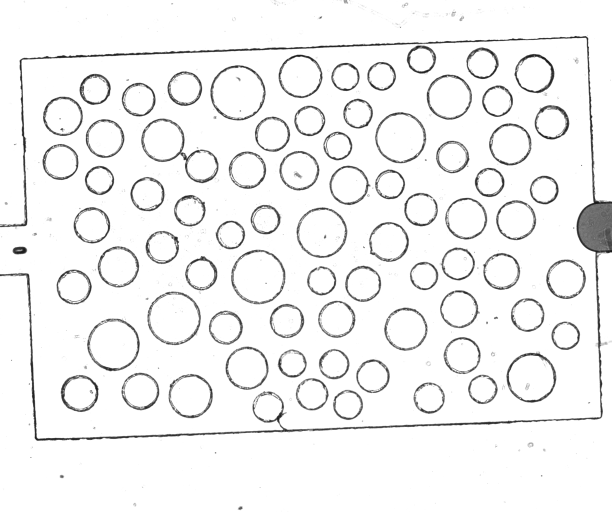}%
    \hfill%
    \includegraphics[width=0.16\columnwidth]{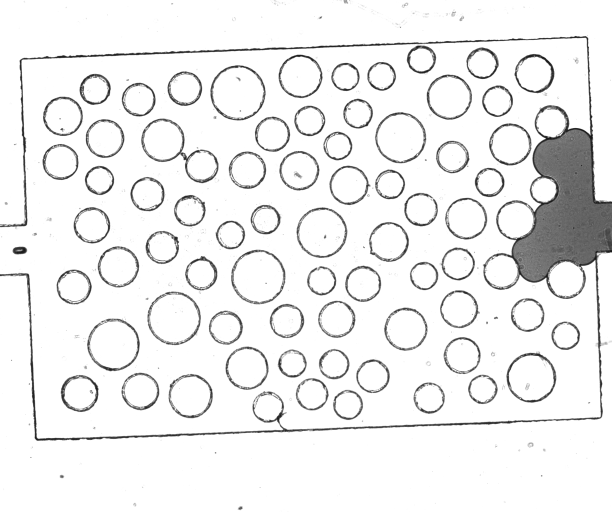}%
    \hfill%
    \includegraphics[width=0.16\columnwidth]{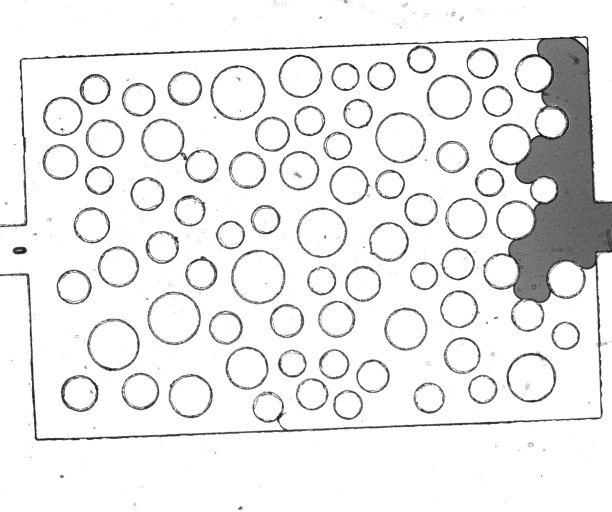}%
    \hfill%
    \includegraphics[width=0.16\columnwidth]{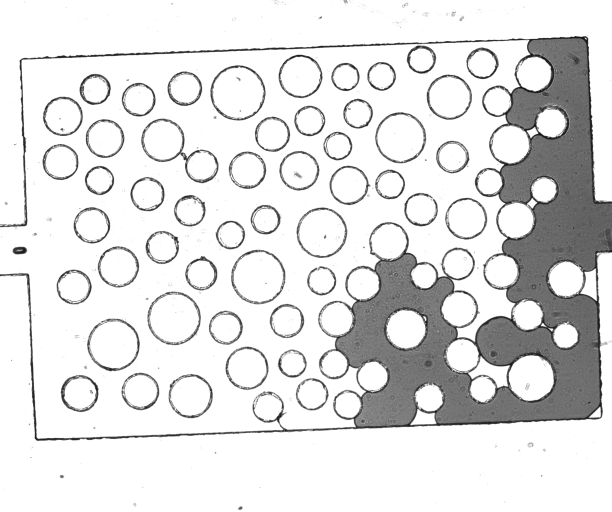}%
    \hfill%
    \includegraphics[width=0.16\columnwidth]{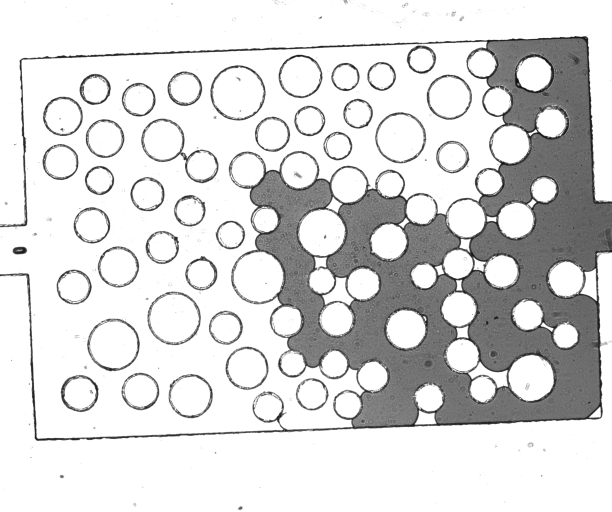}%
    \hfill%
    \includegraphics[width=0.16\columnwidth]{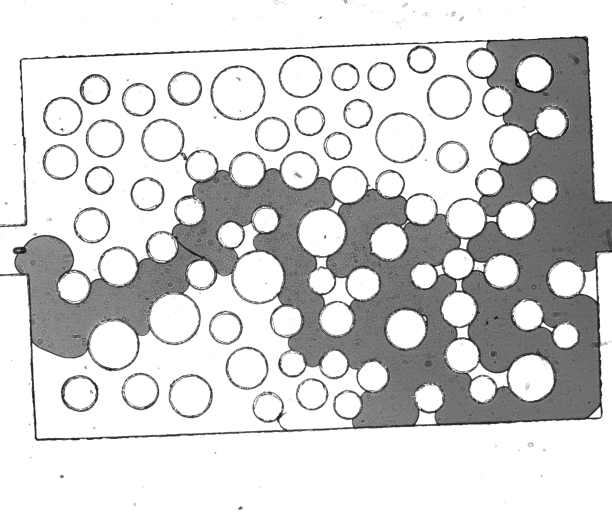}%
  \end{minipage}%
  \\%
  \begin{minipage}[b][\labelHeight][t]{0.04\columnwidth}%
    \subfloat[\label{fig:datasets-octagonal}]{ }%
  \end{minipage}%
  \begin{minipage}{0.96\columnwidth}%
    \includegraphics[width=0.16\columnwidth]{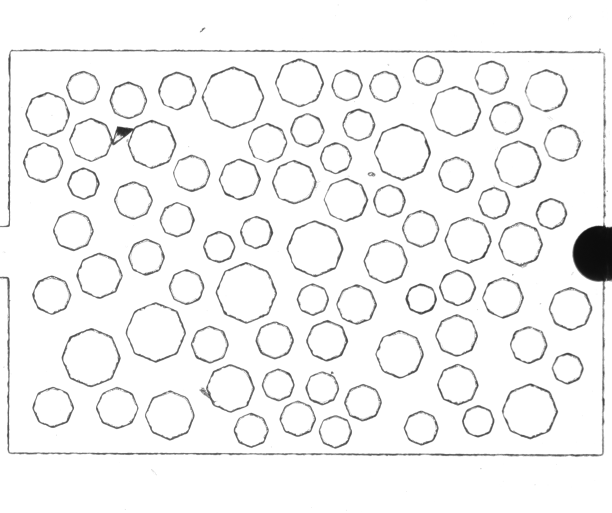}%
    \hfill%
    \includegraphics[width=0.16\columnwidth]{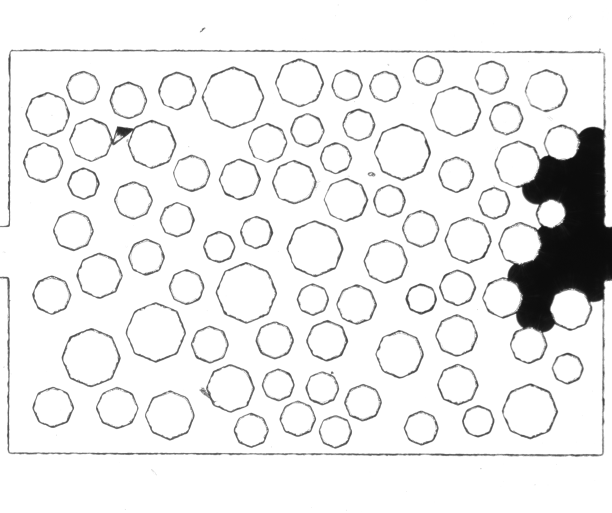}%
    \hfill%
    \includegraphics[width=0.16\columnwidth]{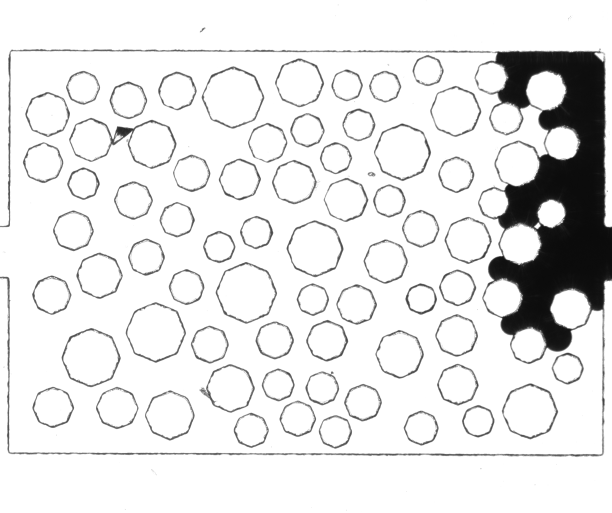}%
    \hfill%
    \includegraphics[width=0.16\columnwidth]{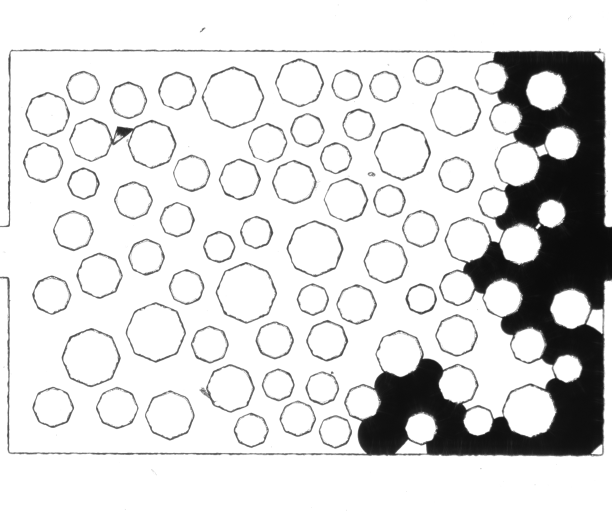}%
    \hfill%
    \includegraphics[width=0.16\columnwidth]{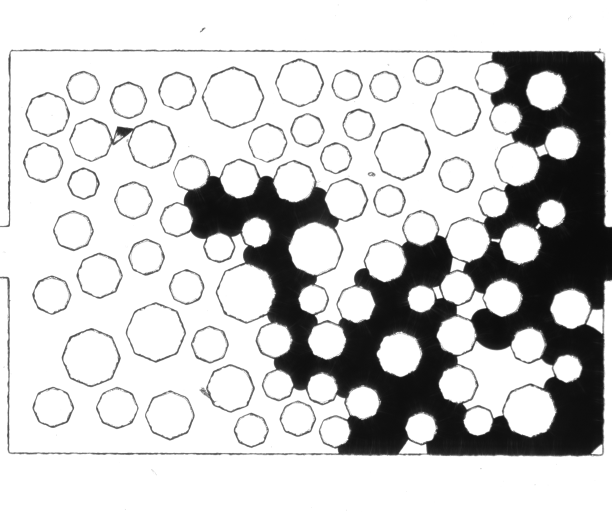}%
    \hfill%
    \includegraphics[width=0.16\columnwidth]{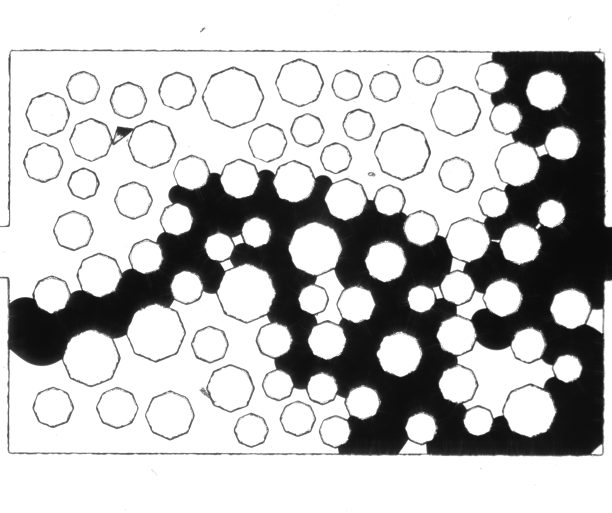}%
  \end{minipage}%
  \\%
  \begin{minipage}[b][\labelHeight][t]{0.04\columnwidth}%
    \subfloat[\label{fig:datasets-triangular}]{ }%
  \end{minipage}%
  \begin{minipage}{0.96\columnwidth}%
    \includegraphics[width=0.16\columnwidth]{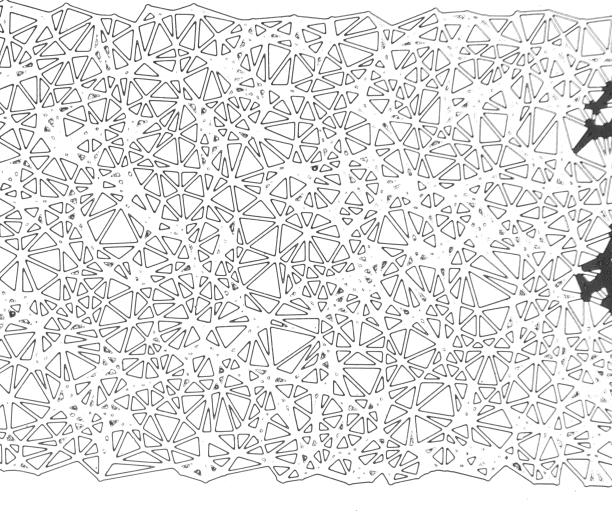}%
    \hfill%
    \includegraphics[width=0.16\columnwidth]{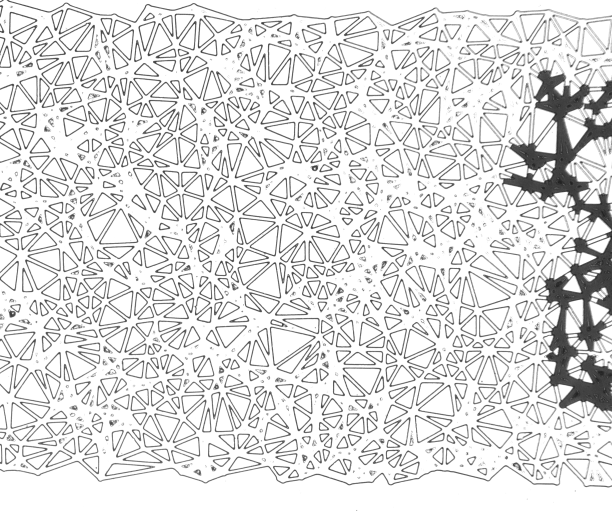}%
    \hfill%
    \includegraphics[width=0.16\columnwidth]{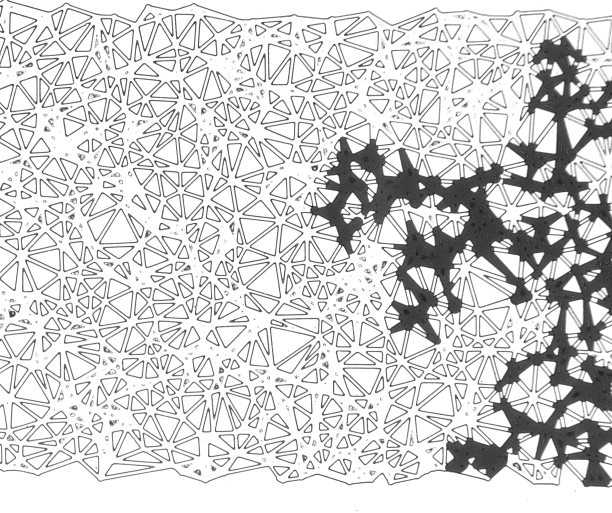}%
    \hfill%
    \includegraphics[width=0.16\columnwidth]{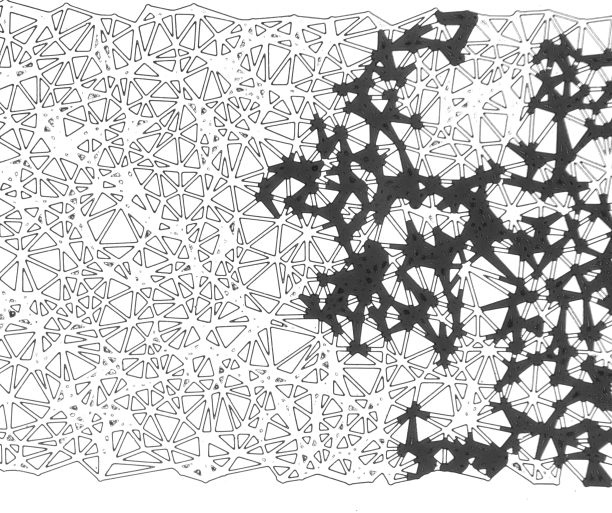}%
    \hfill%
    \includegraphics[width=0.16\columnwidth]{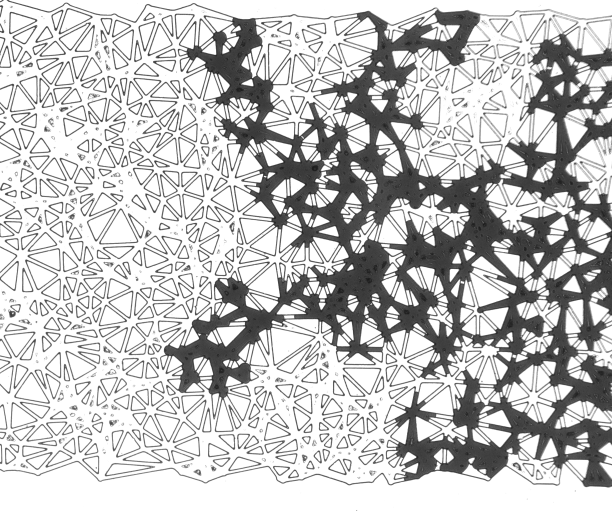}%
    \hfill%
    \includegraphics[width=0.16\columnwidth]{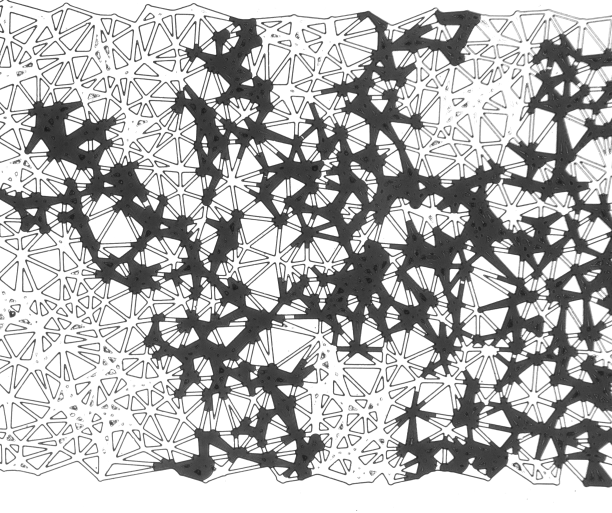}%
  \end{minipage}%
  \vspace{4pt}%
  \caption{%
    Overview of our datasets with different solid geometries: \subref{fig:datasets-circular}~circular, \subref{fig:datasets-octagonal}~octagonal, and \subref{fig:datasets-triangular}~triangular.
    The invading fluid~(gray/black) is entering from the right and displaces the defending fluid~(white).
  }%
  \label{fig:datasets}%
\end{figure}

The input to our method is an image series that generally consists of hundreds of grayscale images, i.e., each image can be described as a function of intensity $I^G: (i,j) \mapsto [0,1] \in \mathbb{R}$ for pixel positions \smash{$(i,j) \in \mathbb{N}^2$}.
From these images, we need to distinguish three components: solid structure, defending fluid, and invading fluid.
In our experiments, white, i.e., $I^G(i,j) \approx 1$, always represents the defending fluid (or void space).
On the other hand, the invading fluid and the outline of the solid structures have similar, non-discernible dark gray values, i.e., $I^G(i,j) \approx 0$.
Representative time steps of our datasets can be seen in~\autoref{fig:datasets}.
To ensure that the defending fluid on one side and the invading fluid and solid structures on the other side are classified correctly, we first apply thresholding with user-defined intensity threshold $\segmThres$.
This definition allows us to react to variations in the datasets, which result from measurement uncertainty, e.g., when using different cameras or different settings, such as exposure time and frame rates.
An example where a high threshold is needed can be seen in~\autoref{fig:datasets-circular}.
Here, the light gray values of the invading fluid would otherwise be classified as void space, i.e., as defending fluid.
Note that we denote the image for discrete frame $\tau$ of the input image series as $I^\tau(i,j)$.

Errors from measurement can also be observed in the form of noise, which can manifest both spatially and temporally.
Common approaches, such as applying Gaussian blur, would lead to grayscale values between the intended black-and-white values, which lead to wrongly segmented pixels.
Therefore, we handle noise only later in the pipeline, where we can apply more knowledge to the denoising process.
Even though some of the herein presented datasets exhibit massive noise, our approach usually still provides valuable results~(\autoref{fig:graph-ensemble-4-10} and \ref{fig:graph-ensemble-4-10-simplified}).

In the following, we detail our method, starting with condensing our input image series to a single image~(\autoref{sec:displacement-map}).
Although this is a preprocessing step~(\autoref{fig:time-triangular}), domain scientists can already leverage the shown velocity frequencies for analysis.
From this image, we then derive the notion of ``flow fronts''~(\autoref{sec:flow-front}) and assign each a unique label~(\autoref{sec:flow-segmentation}) used to create a graph representation of the displacement process~(\autoref{sec:graph}).
While the displacement graph in its spatial embedding already provides insights~(\autoref{fig:graph-triangular-simplified}), different layouts may be of interest to domain scientists to analyze certain processes~(\autoref{fig:graph-triangular-mainchannel}).

\subsection{Flow Tracing}

The goal is to process the input image series in such a way that allows the creation of a graph that fully covers the displacement process.
For this, we assume that once flooded regions of the void space stay flooded, i.e., it does not have to be considered that the invading fluid itself is displaced.
\revision{This assumption holds for the scenario under investigation, where capillary forces govern the flow.}

Our approach contains two steps that build upon each other.
First, the time domain is projected to values in a time map (\autoref{sec:displacement-map}).
Thus, we get only one resulting image from the input image series.
Using this image, we can define \emph{flow fronts} as the displacement that occurred within a single time frame~(\autoref{sec:flow-front}).
Second, applying this definition to the time map, we assign each flow front a unique value~(label) and compute metrics that are representative for this flow front~(\autoref{sec:flow-segmentation}).

\subsubsection{Time Map}
\label{sec:displacement-map}

As mentioned, we assume that areas once flooded by the invading fluid stay flooded, which holds for the capillary flow that is of main interest for our analysis.
Crucially, this allows us to condense the time domain to a single value per pixel.
This value then indicates the time in which the invading fluid first covered the pixel.
Additionally, we store a special value if the pixel belongs to a solid structure or was never covered by the invading fluid.
Note that we discuss the limitations of this particular approach in~\autoref{sec:discussion}.

The algorithm starts by initializing the time map $M^0(i,j)$ using the segmented first input image $I^0(i,j)$ in the series as
\begin{equation}
    M^0(i,j) = \left\lbrace \MatrixX{0, & I^0(i,j)=0 \\ \infty, & \text{else}} \right.,
\end{equation}
assuming that the invading fluid is not present in the first time step, yet~(which is generally the case for our data).
Subsequently, for each image $I^\tau$ with discrete time values $0 < \tau \leq T$, the time map is updated:
\begin{equation}
    M^\tau(i,j) = \left\lbrace \MatrixX{ \min\lbrace M^{\tau-1}(i,j), t \rbrace, & I^\tau(i,j)=0 \\ M^{\tau-1}(i,j), & \text{else}}\right..
\end{equation}
The resulting final time map $M(i,j) := M^T(i,j)$ then stores $0$ if the pixel represents the solid structure, $\infty$ if the defending fluid was not displaced, and $\tau$ if the invading fluid first filled the pixel in time step $\tau$.
This allows us to visualize the time map to show the progression of the invading fluid.
Additionally, we get velocity information by applying a periodic color map to the discrete time values~(\autoref{fig:time-triangular}).
One period thus always relates to the same physical time within a dataset, which allows us to easily distinguish between areas of low or high velocities, and hence to observe velocity jumps.
Additionally, we allow the user to select a specific time frame for highlighting, thus identifying all areas that pertain to the selected time value~(red areas in \autoref{fig:displacement-overview}).

\subsubsection{Flow Front}
\label{sec:flow-front}

From our time map $M$, we can identify connected regions of pixels with the same time value.
We call such a region a flow front, as it represents the movement of the invading fluid in the respective time step.
This will provide the basis for generating our displacement graph.
A flow front can be defined by iteratively expanding an initial region $\mathcal{F}^0(i,j) = \left\lbrace (i,j) \right\rbrace$ by adding the time-local pixel neighborhood
$\mathcal{N}(i,j) = \left\lbrace (k,l) ~|~ M(i,j) = M(k,l) \land~ (i-1 \leq k \leq i+1) ~\land~ (j-1 \leq l \leq j+1) \right\rbrace$
of all its already contained pixels, thus
\begin{equation}
  \mathcal{F}^{\eta+1} = \bigcup_{(i,j)\in\mathcal{F}^{\eta}}~ \mathcal{N}(i,j).
\end{equation}
Eventually, this converges to a flow front $\mathcal{F} := \mathcal{F}^N$, when no pixels can be added anymore, i.e., $\mathcal{F}^N = \mathcal{F}^{N+1}$.

\begin{figure}[t]%
  \centering%
  \subfloat[\label{fig:displacement-overview}]{%
    \begin{tikzpicture}%
      \node[anchor=south west, inner sep=0] (image) at (0,0) {\includegraphics[width=0.8\columnwidth]{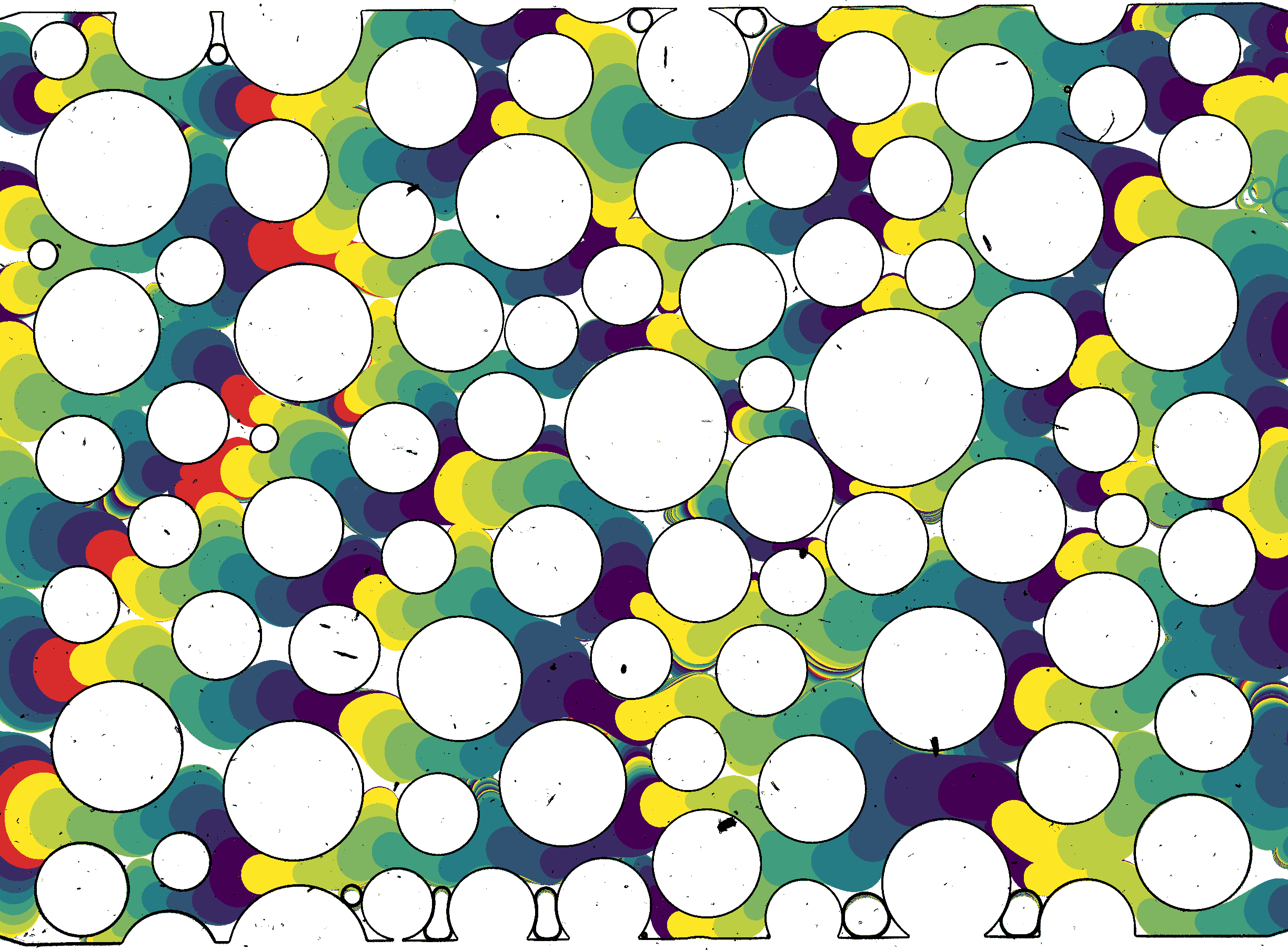}};%
      \draw[red,very thick] (3.52,2.24) rectangle (4.0,2.56);
    \end{tikzpicture}%
  }%
  \\%
  \subfloat[\label{fig:displacement-original}]{%
    \vspace{-0.8em}%
    \includegraphics[width=0.49\columnwidth,trim={0cm 1cm 0cm 4cm},clip]{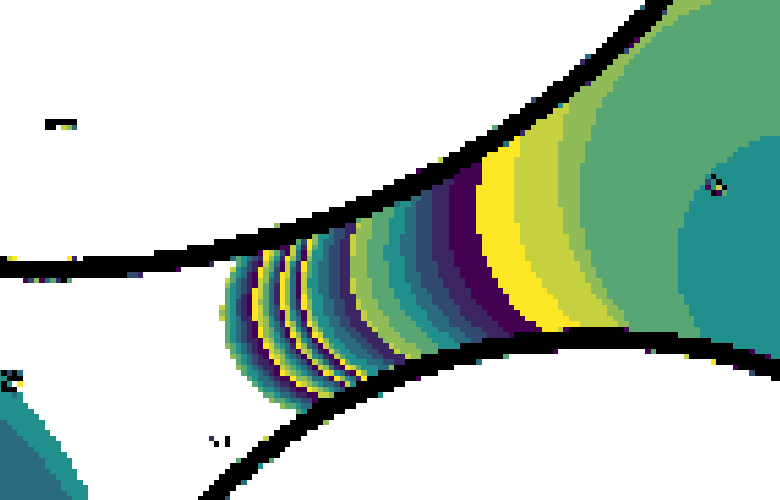}%
  }%
  \hfill%
  \subfloat[\label{fig:displacement-quantized}]{%
    \vspace{-0.8em}%
    \includegraphics[width=0.49\columnwidth,trim={0cm 1cm 0cm 4cm},clip]{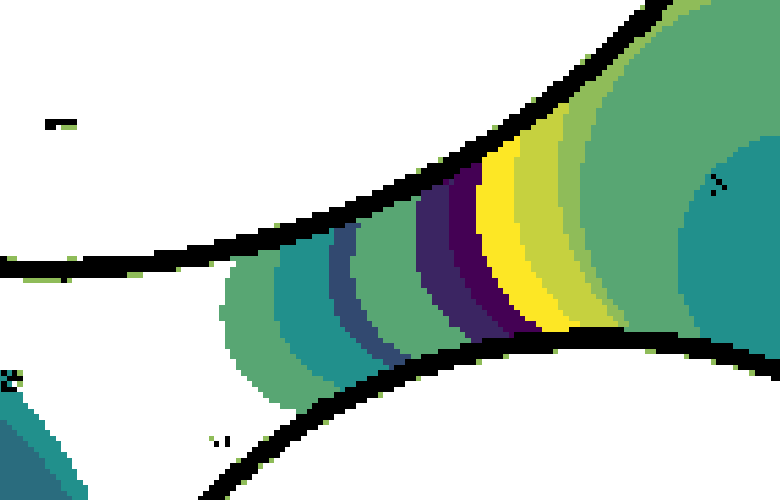}%
  }%
  \caption{Time map with discrete time steps mapped to a periodic color map.
    \subref{fig:displacement-overview}~Time map for the whole domain~(quantized).
    \subref{fig:displacement-original}~Zoom-in on a region~(red box) where small velocities lead to sub-pixel flow front propagation.
    \subref{fig:displacement-quantized}~Quantization ``resolves'' this issue and removes noise.
  }%
  \label{fig:displacement}%
\end{figure}

\textbf{Combination of Small Flow Fronts (\& Noise Reduction).}
While small velocities pose no problems when visualizing the time map, sub-pixel flow propagation leads to an extensive amount of separate flow fronts being detected.
See, for example, the regions with high frequency in~\autoref{fig:displacement-original}.
As we want to use these flow fronts as a basis for graph generation, i.e., generate one node per flow front, this would lead to a massive amount of unwanted nodes being created that do not correctly represent the fluid flow.
Thus, in such a case, we want to combine neighboring areas of similar time values, as indicated in~\autoref{fig:displacement-quantized}.
Although this means that we lose information in regions of high frequency, analogous to applying a low-pass filter, it allows us to generate a flow graph that now correctly represents the flow in terms of topology.
To this end, we apply a restricted quantization, adjusting the time value of small flow fronts.
Thus, different flow fronts can be combined into larger fronts, as they now exhibit the same time value.
Because we do not adjust large areas, we have to restrict the quantization to not lose the invariance of monotonicity of time between adjacent areas.
This has to be done in such a way that small flow fronts are not assigned values smaller or larger than respective large neighboring areas.
To define a \emph{small area}, we have to introduce a threshold $\gamma$, which generally depends on the throat size within the porous medium because the combined area should be able to at least span the whole width of a throat.
In all our cases, a value of $\gamma = 100$ was selected.
\revision{In general, this value should be a small multiple of the maximum throat size of the domain in pixels.
Else, smaller values would not combine enough areas.
In contrast, very large values might affect the topology of the graph.
When the above considerations are taken into account, our quantization approach is very robust to changes in its parameter.}

Let now $M_{min}(\mathcal{F})$ be the minimum time of a neighbor $\mathcal{F}_{min}$ of $\mathcal{F}$, whose size is above the defined threshold, i.e., $A(\mathcal{F}_{min}) \geq \gamma$.
If such a neighbor does not exist, set $M_{min}(\mathcal{F})$ to $-\infty$.
Now, iteratively adjust the time value of all small flow fronts until their size exceeds the threshold, or their time value is reduced to zero (we then set the time value to $\infty$).
We do this by quantizing in each iteration step an increasing range of values $2^i$, starting at $i=1$:
\begin{enumerate}
  \item Set range $r=2^i$ and calculate $\lambda = M_{old}(\mathcal{F})\mod r$
  \item The new time is the quantized time value $M_{new}(\mathcal{F})^* = M_{old}(\mathcal{F}) - \lambda.$
  \item As we always only lower the time value, we have to restrict the new time value to stay within the bounds enforced by $M_{min}(\mathcal{F})$ by clamping the new value: $
      M_{new}(\mathcal{F}) = \max\left\lbrace M_{new}(\mathcal{F})^*, M_{min}(\mathcal{F})\right\rbrace$
  \item Perform flood fill only on the adjusted flow fronts.
    As this will propagate also into areas that have not been adjusted, areas of now the same time value will merge and form larger flow fronts.
\end{enumerate}
Please note that we quantize using a modulo approach.
This ensures that the number of values, except for large enough flow fronts, is halved in comparison to the previous iteration.
Thus, it is guaranteed that flow fronts will be combined with either other small flow fronts that are being adjusted to the same value, or, because of the restriction, with neighboring large areas.
As a side effect, not only flow fronts are combined with this approach, but also noise is drastically reduced.

\subsubsection{Flow Segmentation}
\label{sec:flow-segmentation}

Now, only one step is missing before we can generate our graph that detects topological flow events, such as splits and merges of fingers.
For this, we assign each individual flow front $\mathcal{F}$ a unique label.
Additionally, we calculate and store for each flow front the following information and derived quantities that we call metrics:
\begin{itemize}
  \item \textbf{frame time} $M$,
  \item \textbf{area}, as number of pixels $(i,j) \in \mathcal{F}$,
  \item \textbf{center of mass} $\vec{p} = \frac{1}{|\mathcal{F}|} \sum_{(i,j)\in\mathcal{F}} (i~~j)^{-1}$,
  \item \textbf{fluid-fluid interface length} as number of pixels $(i,j)$ that have a neighbor $(k,l)$ with $M(k,l) > M(i,j)$,
  \item \textbf{fluid-solid interface length} as number of pixels $(i,j)$ that have a neighbor $(k,l)$ with $M(k,l) = 0$,
  \item \textbf{bounding rectangle} of $\mathcal{F}$, and
  \item \textbf{velocity magnitude} $||\vec{v}||$ as the Hausdorff distance between subsequent fluid-fluid interfaces of the flow front~(\autoref{eq:hausdorff}).
\end{itemize}
\revision{These metrics can later be mapped to visual channels of our graph or they can be visualized as charts in our interactive tool~(cf.~\autoref{sec:results}).}

\subsection{Graph}
\label{sec:graph}

\begin{figure}[t]%
  \centering%
  \begin{tikzpicture}%
    \node[anchor=south west, inner sep=0] (image) at (0,0) {\includegraphics[width=0.37\columnwidth]{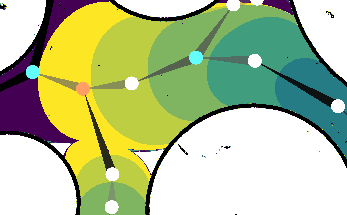}};%
    \node[anchor=south west] at (0.1,-0.3) {\subfloat[\label{fig:edges-merge}]{}};%
  \end{tikzpicture}%
  \hspace{0.5em}%
  \begin{tikzpicture}%
    \node[anchor=south west, inner sep=0] (image) at (0,0) {\includegraphics[width=0.37\columnwidth]{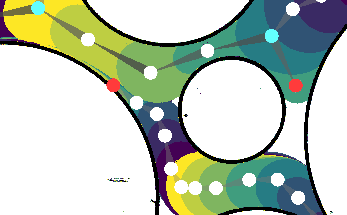}};%
    \node[anchor=south west] at (0.1,-0.3) {\subfloat[\label{fig:edges-deadend}]{}};%
  \end{tikzpicture}%
  \caption{%
    Edge creation during graph generation.
    \subref{fig:edges-merge}~If two flow fronts of the same time frame arrive at a junction, they merge into a single flow front.
    This is reflected in the graph by a node with two incoming edges~(orange).
    \subref{fig:edges-deadend}~If at a junction two flow fronts of different time frames meet, the flow already progressed through the first flow front, and the later arriving fluid goes into a ``dead end''.
    This is reflected as a sink node~(red).
  }%
  \label{fig:edges}%
\end{figure}

The creation of the nodes $\mathcal{N}$ for the displacement graph $G=\lbrace \mathcal{N}, E \rbrace$ is now simple.
For every unique flow front, we create a node with position at the respective center of mass.
Additionally, we store the metrics calculated in~\autoref{sec:flow-segmentation} at the nodes for later visualization.
To complete the graph, we now only have to connect nodes of adjacent flow fronts.
However, edges $E$ should only be created between two nodes $\mathcal{N}_i$ and $\mathcal{N}_j$, with $M(\mathcal{N}_i)<M(\mathcal{N}_j)$, if $M(\mathcal{N}_i)$ is the local maximum of the neighbors of $\mathcal{N}_j$ below $M(\mathcal{N}_j)$.
The reason for this decision is that we do not want to connect to a node that arrives ``late''---and thus goes into a ``dead end''.
An example of this behavior shown in~\autoref{fig:edges}.

\begin{figure}[t]%
  \centering%
  \subfloat[\label{fig:hausdorff-issue}]{%
    \includegraphics[height=0.3\columnwidth]{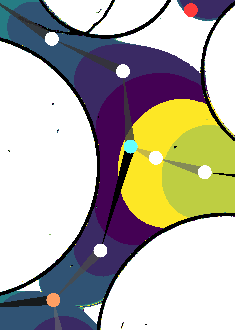}%
  }%
  \hspace{0.5em}%
  \subfloat[\label{fig:hausdorff-solution}]{%
    \includegraphics[height=0.3\columnwidth,trim={2cm 2.5cm 2cm 1cm},clip]{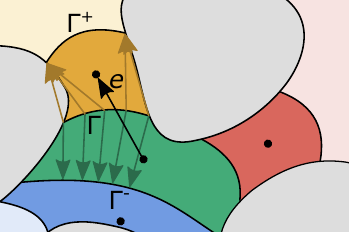}%
  }%
  \caption{%
    Velocity calculation at the graph edges.
    \subref{fig:hausdorff-issue}~At splits with wide angle, the large distance between the centers of mass is not representative when calculating velocity.
    \subref{fig:hausdorff-solution}~Modified Hausdorff distances for the calculation of the velocity for the specified edge~(black arrow).
    Only minimum distances from the interface $\Gamma$ are considered in forward~(orange arrows) and reverse time~(green arrows), respectively.
  }%
  \label{fig:hausdorff}%
\end{figure}

Similar to the metrics stored at the nodes, we want to store the velocity at the edges.
As an intuitive approach, the first idea was to simply use the distance between the centers of mass.
This, however, proved to yield inaccurate results at junctions, as the centers of mass can be far apart, although the distance for the fluid is obviously rather short, as can be seen for an example in~\autoref{fig:hausdorff-issue}.
Thus, we need a quantity that better describes the movement of the fluid along the edge.
For this, we derive the velocity from a modified Hausdorff distance.
As the Hausdorff distance describes the distance between two shapes, i.e., interfaces, it can easily define the displacement between fluid-fluid interfaces $\Gamma^-,\Gamma^+$ of two subsequent time steps:
\begin{equation}
    \label{eq:hausdorff}
    d_H(\Gamma^-,\Gamma^+) = \max\left\lbrace \max_{p \in \Gamma^-} d(p, \Gamma^+), \max_{q \in \Gamma^+} d(q, \Gamma^-) \right\rbrace
\end{equation}
with $d(x,\Gamma) = \min_{y \in \Gamma} d(x,y)$ and Euclidean distance $d(\cdot,\cdot)$.
However, we want to obtain the velocity that describes the movement of the fluid along a specific edge.
This means that the Hausdorff distance should be calculated only with respect to the fluid-fluid interface that is described by this edge, as illustrated in~\autoref{fig:hausdorff-solution}.
Hence, the idea is to calculate the \emph{forward}~(orange arrows) and \emph{backward}~(green arrows) Hausdorff distances, i.e., we are interested in the maximum of the minimum distances from the specified fluid-fluid interface $\Gamma$ to the fluid-fluid interface in either forward~($\Gamma^+$) or reverse~($\Gamma^-$) time.
Hence, we define forward Hausdorff distance $d_H^+$ and backward distance $d_H^-$ as
\begin{equation}
  \begin{split}
    d_H^\pm(\Gamma,\Gamma^\pm) := \max_{p \in \Gamma} d(p,\Gamma^\pm).
  \end{split}
\end{equation}
The velocity at the edge is thus simply the average of the forward~($d_H^+$) and backward~($d_H^-$) Hausdorff distances, weighted by the time difference between the two involved flow fronts.

\begin{figure}[t]%
  \centering%
  \begin{tikzpicture}%
    \node[anchor=south west, inner sep=0] (image) at (0,0) {\includegraphics[width=0.245\columnwidth,trim={2cm 2.5cm 6.3cm 0cm},clip]{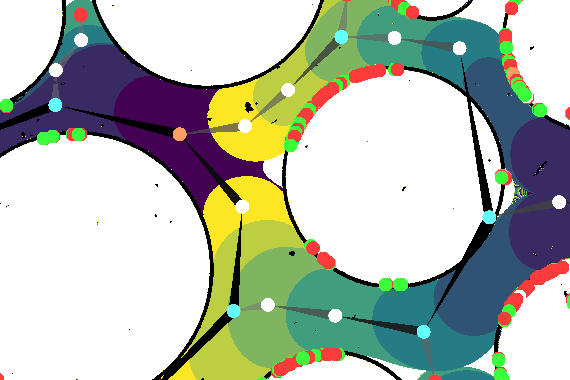}};%
    \node[anchor=south west] at (0.1,-0.3) {\subfloat[\label{fig:graph-original}]{}};%
  \end{tikzpicture}%
  \hfill%
  \begin{tikzpicture}%
    \node[anchor=south west, inner sep=0] (image) at (0,0) {\includegraphics[width=0.245\columnwidth,trim={2cm 2.5cm 6.3cm 0cm},clip]{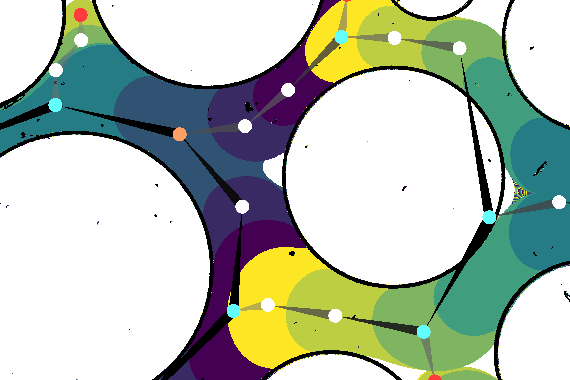}};%
    \node[anchor=south west] at (0.1,-0.3) {\subfloat[\label{fig:graph-fixed}]{}};%
  \end{tikzpicture}%
  \hfill%
  \begin{tikzpicture}%
    \node[anchor=south west, inner sep=0] (image) at (0,0) {\includegraphics[width=0.245\columnwidth,trim={2cm 2.5cm 6.3cm 0cm},clip]{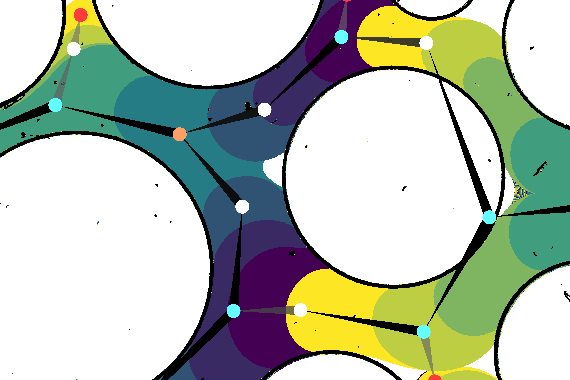}};%
    \node[anchor=south west] at (0.1,-0.3) {\subfloat[\label{fig:graph-combined}]{}};%
  \end{tikzpicture}%
  \hfill%
  \begin{tikzpicture}%
    \node[anchor=south west, inner sep=0] (image) at (0,0) {\includegraphics[width=0.245\columnwidth,trim={2cm 2.5cm 6.3cm 0cm},clip]{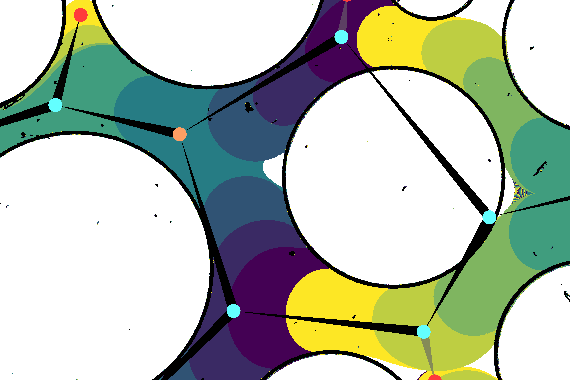}};%
    \node[anchor=south west] at (0.1,-0.3) {\subfloat[\label{fig:graph-removed}]{}};%
  \end{tikzpicture}%
  \caption{%
    Time map with the displacement graph at different stages of dealing with noise, and simplifications.
    \subref{fig:graph-original}~Graph before applying any fixes.
    Thus, noise around the solid phase is still present.
    Graph after \subref{fig:graph-fixed}~applying all fixes, \subref{fig:graph-combined}~combining all nodes with only one incoming and one outgoing edge into one representative node, and \subref{fig:graph-removed}~removing all these nodes to provide the fully simplified graph.
  }%
  \label{fig:graph-stages}%
\end{figure}

\textbf{Dealing with Noise.}
The resulting graph is still strongly influenced by noise~(\autoref{fig:graph-original}).
Fortunately, most of it can easily be dealt with by applying a few fixes that are guaranteed to preserve the correct topology of the graph and only remove parts that result from noise.
The first of these fixes removes isolated nodes.
These do not contribute to the graph and do not provide any information.
As a second step, we remove source nodes that are not in a region where sources are expected.
These should only be present at the inlet of the porous medium.
In all our datasets, this is at the right side of the domain, but it can be defined by the user.
After removing these sources, we can iteratively remove all nodes that become sources through this step.
This gets rid of a lot of noise, especially in the neighborhood of solid structures.
These artifacts are caused by small movements of the substrate during the experiments or result from poor image quality.
As a last step, we remove sink nodes that have neighbors with larger time frame, as these cannot be real sinks, i.e., at a later time, the fluid keeps moving and flow is thus guaranteed to continue from its neighbors.
See a fixed example in~\autoref{fig:graph-fixed}.

\textbf{Simplification.}
Until now, we only applied modifications to the graph for its improvement.
The current graph can thus be already applied for visualization and to store metrics for later analysis.
However, we want to apply simplifications to the graph in order to reduce its complexity.
These simplifications should, in general, maintain the topology of the fluid flow through the porous network.
\revision{The idea is therefore to remove nodes that do not provide any information regarding topology, i.e., nodes that only have one incoming and one outgoing edge.}
This allows us to abstractly visualize the graph, either in its spatial embedding or by applying layout algorithms, and thus compare graphs of different experiments in an ensemble dataset.
We do this in two steps: we first combine subsequent nodes that fit this criterion into one node~(\autoref{fig:graph-combined}).
This node can still contain information about the movement of the fluid, such as the area covered by the node and channel velocity, and can function as an intermediate representation.
In the second step, all combined nodes are removed~(\autoref{fig:graph-removed}).
Here, however, it is not possible to accurately maintain the metrics stored at the nodes, as parts of the removed nodes would need to be added to the connecting nodes on either side.
Thus, the visualized properties are extracted from the non-simplified graph in order to not average over the metrics.
Additionally, we potentially retain some of the trivial nodes, e.g., if their time frame coincides with breakthrough time, where domain scientists are interested in the distribution of flow and the shape of the graph.
Further, we provide the option to preserve nodes at velocity jumps~(\autoref{fig:graph-octagonal-jumps}) to more accurately map velocity to the edges of the graph and to investigate Haines jumps~(\autoref{sec:results-domain}).
These jumps are detected by looking at the velocities at neighboring nodes and imposing a threshold on the velocity ratio, i.e., on acceleration.
\autoref{fig:graph-triangular-simplified} shows an example of our simplified graph, called \emph{displacement graph}, with node positions in their spatial embedding.
In addition to node positions, we also map quantities stored at the nodes to visual channels: the area of the respective flow front denotes the size of the nodes, and color depicts time~(white to red) to emphasize the temporal aspect of the displacement process.
The directed edges of the graph are represented by arrows that indicate flow direction, and edge width conveys velocity information, with larger velocities mapped to wider edges.

\textbf{Breakthrough Graph.}
As already discussed, source nodes can be restricted to a region at the inlet, as specified for each dataset by the user.
Similarly, outflow is defined as flow fronts reaching the area of the outlet, which usually is on the opposite side from the inlet.
The first occurrence of such an outflow event is called the \emph{breakthrough}, which means that the invading fluid now connects inlet and outlet.
This point usually marks a general change in flow behavior.
Thus, identifying breakthrough and the time frame associated with it is of interest to domain experts.
In~\autoref{fig:displacement-overview}, all flow fronts at the time of breakthrough are highlighted in red, with breakthrough occurring in the lower left part of the image.
To analyze the flow leading up to breakthrough, we extract the ``main channel'' that connects the breakthrough node in the graph to one of the source nodes at the inlet.
In the case that this main channel is not trivially defined, i.e., there exists more than one path in the graph that connects these two nodes, we take the path with the largest area, i.e., where most of the fluid went through.
In the triangular dataset the main channel can be uniquely identified (\autoref{fig:graph-triangular-mainchannel}).

The importance of breakthrough and of the main channel motivates us to not only show the simplified graph in its spatial embedding, with centers of mass mapped to node positions, but also to consider a different layout to highlight the main channel and its branches.
To this end, similar to Eulzer et al.~\cite{Eulzer2021}, we adjust the node positions of the main channel to lie on a horizontal line, while preserving the relative distances between them.
All other nodes are then re-positioned by applying a force-based layout algorithm.
For this, we use \emph{ForceAtlas~2}~\cite{Jacomy2014} as it provides good visual results (see~\autoref{fig:graph-triangular-mainchannel}).
While also considering other layout algorithms, such as \emph{Yifan Hu}~\cite{Hu2006} for the preservation of distances, non-unique main channels and non-tree graphs do not allow to preserve relative distances.

\section{Implementation and Interactive Framework}
\label{sec:visualization}

The previously described steps are all implemented in MegaMol~\cite{Grottel2015}.
This visualization framework provides a modular approach and ready-to-use renderers.
We therefore added new modules to handle input image series, preprocess the images, and condense them into a single time map $M$.
Another module then uses the input time map to generate our graphs, which can be rendered or exported to file.
This process can be influenced by the user by adjusting parameters.
\revision{As these steps are all performed on images, the complexity for creating the time map is in $\text{O}(T \cdot N)$, the following steps are in $\text{O}(N)$, where $T$ is the number of input images, i.e., time steps, and $N$ is the number of pixels.
For the investigated datasets, we measured between $8.53 s$ and $61.88 s$ from loading the datasets until rendering of the results.
Please see the supplemental material in the Appendix for more details.}
For the layout of the breakthrough graph and its visualization, we used the exported graph files for use in Gephi.
Thus, we could apply readily available algorithms to apply a force-based layout to the nodes that are not part of the main channel.
Additionally, we mapped the properties calculated in MegaMol to visual channels of the nodes and edges of the graph.

\begin{figure}[t]%
  \centering%
  \begin{tikzpicture}%
    \node[anchor=south west, inner sep=0] (image) at (0,0) {\includegraphics[width=0.99\columnwidth,trim={2cm 2.5cm 4.3cm 0cm},clip]{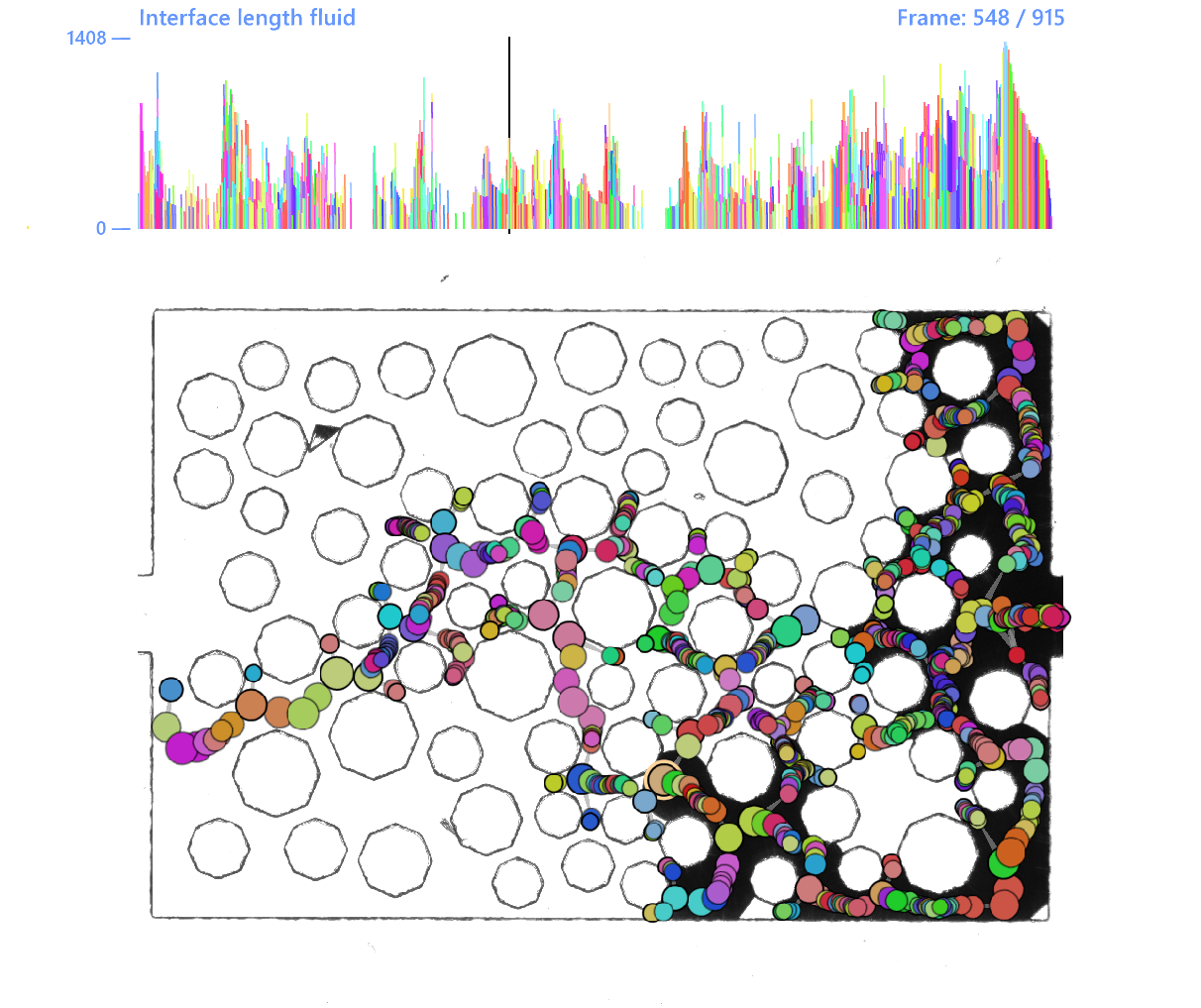}};%
    \draw[ggreen,very thick] (0,0) rectangle (8.8,5.6);
    \node at (0.4,2.8) {\textcolor{ggreen}{\circledNumber{2}{\One}}};
    \draw[ggreen,very thick] (0,5.8) rectangle (8.8,8.05);
    \node at (0.4,6.9) {\textcolor{ggreen}{\circledNumber{2}{\Two}}};
    \draw[ggreen,very thick] (3.77,5.9) rectangle (3.97,7.8);
    \node at (4.25,7.65) {\textcolor{ggreen}{\circledNumber{2}{\Three}}};%
    \draw[ggreen,very thick] (7.1,7.7) rectangle (8.7,8);
    \node at (6.8,7.75) {\textcolor{ggreen}{\circledNumber{2}{\Four}}};
    \draw[red,very thick] (4.95,1.1) rectangle (5.5,1.65);
  \end{tikzpicture}%
  \vspace{6pt}%
  \caption{%
    Example frame from the experiment with octagonal solid structure in our interactive visual analysis framework \emph{LuaVis}.
    \smash{\circledNumber{1.2}{\One}}:~Non-simplified graph in its spatial embedding with input image of the selected frame in the background.
    \smash{\circledNumber{1.1}{\Two}}:~Stacked bar chart for the selected metric, i.e., fluid-fluid interface length.
    \smash{\circledNumber{1.1}{\Three}}:~Marker in the background for the currently selected time step, and
    \smash{\circledNumber{1.1}{\Four}}:~corresponding frame index.
    Nodes are colored randomly to be relatable to the values of the stacked bars and active nodes of the current frame are marked~(see orange node in red box).
  }%
  \label{fig:framwork}%
\end{figure}

For interactive visual analysis, we further developed a standalone application in Lua~(\emph{LuaVis}) that uses the exported graphs from Mega\-Mol.
This framework allows us to visualize in the background the original input image of the experiment for a user-selected time frame together with the exported graph~(bottom in~\autoref{fig:framwork}) and a visualization of related metrics~(top).
Among these metrics are for the corresponding time step the velocity, the covered area, the lengths of the fluid-fluid and fluid-solid interfaces, and the number of fingers.
\revision{These metrics and their plots are of interest to our domain scientists and have been discussed non-interactively in the paper by Frey et al.~\cite{Frey2021}.}
By interactively selecting time frames from the dataset, the user can jump to interesting frames as guided by the shown metrics in (stacked) bar charts.
The respective nodes of the graph that belong to this time step are then visually highlighted, and the appropriate input image shown as background.
To guide the user to interesting events in the graph, nodes are either colored by \emph{node type}, i.e., according to its incoming and outgoing edges, or are assigned a random color.
The latter can be used to identify the node's influence on the metrics in the stacked bar chart by relating the node color.
Additionally, the user can modify the visualized graph, e.g., hiding parts of the graph that belong to time after the breakthrough, only showing the graph until the selected time frame, and hiding nodes that only have one incoming and one outgoing edge~(graph simplification in~\autoref{sec:graph}).
Further, the shown metrics can be exported in the CSV file format for further analysis.

\revision{The implementation can be found on GitHub: \url{https://github.com/UniStuttgart-VISUS/porous-flow-graph}.}

\section{Results}
\label{sec:results}

In the following, we will discuss the results of our approach on three datasets~\cite{Karadimitriou2023}.
The first one is an ensemble dataset with varying capillary number and viscosity ratio, the second one an ensemble dataset of two different geometry types, and the last one a dataset with a triangular solid structure.
In this section, we focus mostly on the visualization aspects of our method, while the domain experts discuss knowledge gained for porous media research in~\autoref{sec:results-domain}.

\begin{figure*}[t]%
  \centering%
  \def\ensembleSize{0.39}
  \subfloat[\label{fig:time-ensemble-5-0.2} $\ensembleParams{-5}{0.2}$]{%
    \includegraphics[width=\ensembleSize\columnwidth,trim={0cm 1.05cm 0cm 0.75cm},clip]{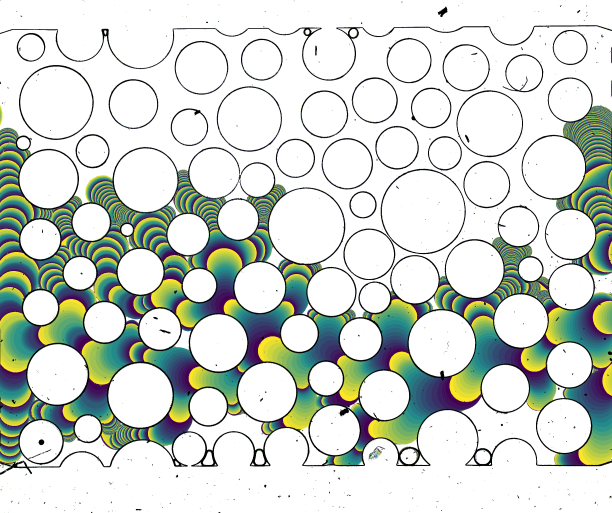}%
  }%
  \hspace{1em}%
  \subfloat[\label{fig:time-ensemble-4-0.2} $\ensembleParams{-4}{0.2}$]{%
    \includegraphics[width=\ensembleSize\columnwidth,trim={0cm 1.3cm 0cm 0.5cm},clip]{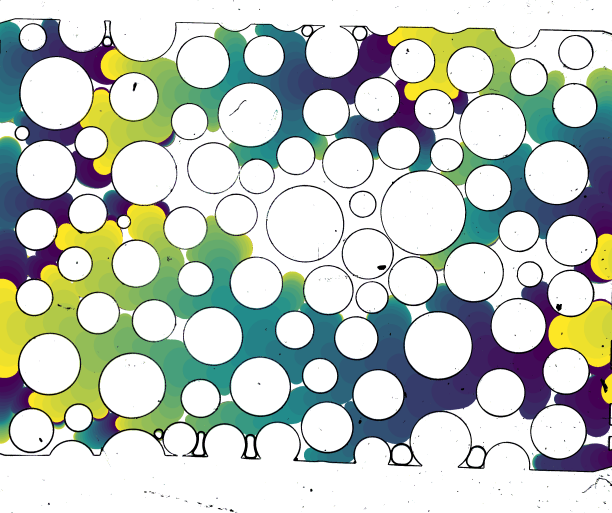}%
  }%
  \hspace{1em}%
  \subfloat[\label{fig:time-ensemble-3-0.2} $\ensembleParams{-3}{0.2}$]{%
    \includegraphics[width=\ensembleSize\columnwidth,trim={0cm 0.9cm 0cm 0.9cm},clip]{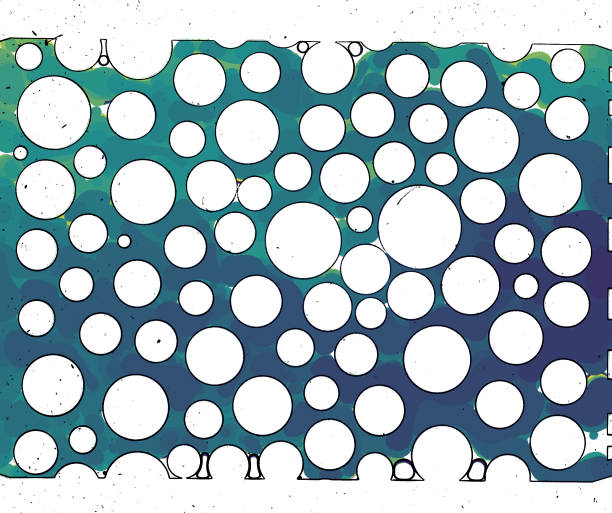}%
  }%
  \hspace{1em}%
  \begin{minipage}{\ensembleSize\columnwidth}{\color{white}.}\end{minipage}%
  \\%
  \subfloat[\label{fig:time-ensemble-5-1} $\ensembleParams{-5}{1}$]{%
    \includegraphics[width=\ensembleSize\columnwidth,trim={0cm 1.3cm 0cm 0.5cm},clip]{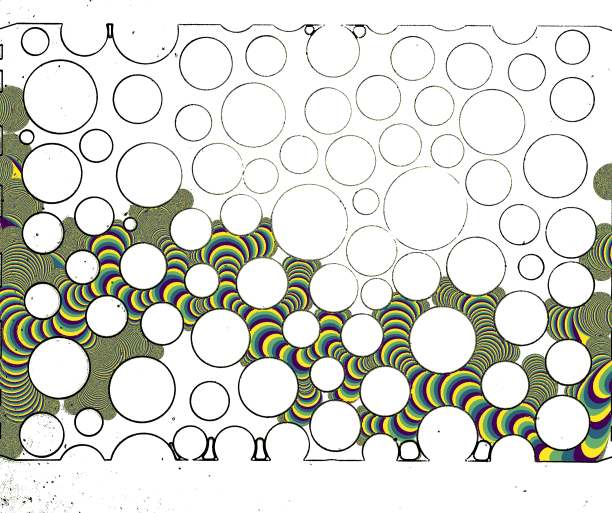}%
  }%
  \hspace{1em}%
  \subfloat[\label{fig:time-ensemble-4-1} $\ensembleParams{-4}{1}$]{%
    \includegraphics[width=\ensembleSize\columnwidth,trim={0cm 1.1cm 0cm 0.7cm},clip]{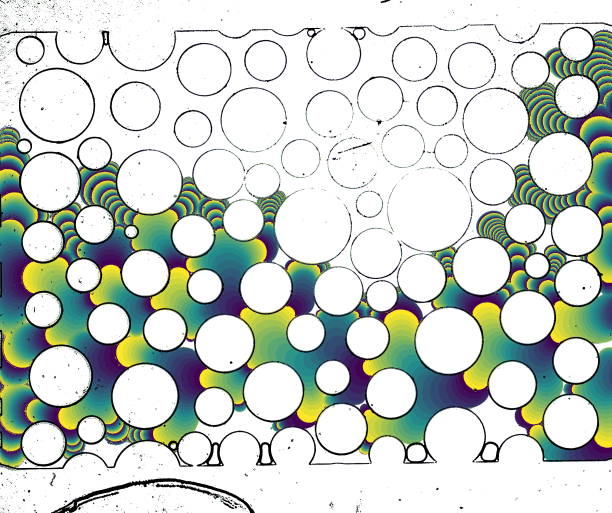}%
  }%
  \hspace{1em}%
  \subfloat[\label{fig:time-ensemble-3-1} $\ensembleParams{-3}{1}$]{%
    \includegraphics[width=\ensembleSize\columnwidth,trim={0cm 0.9cm 0cm 0.9cm},clip]{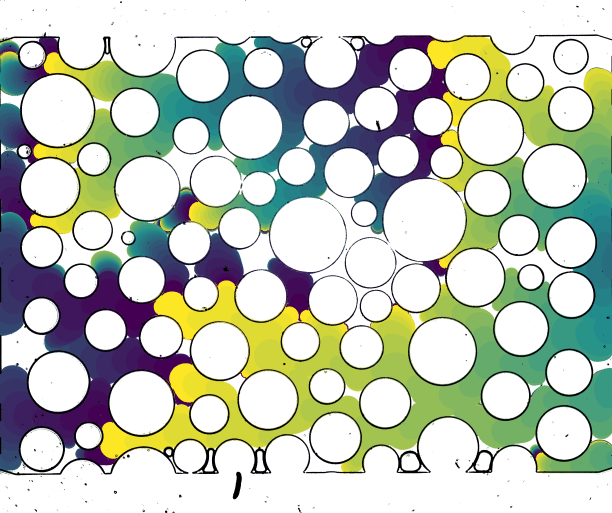}%
  }%
  \hspace{1em}%
  \subfloat[\label{fig:time-ensemble-2-1} $\ensembleParams{-2}{1}$]{%
    \includegraphics[width=\ensembleSize\columnwidth,trim={0cm 0.9cm 0cm 0.9cm},clip]{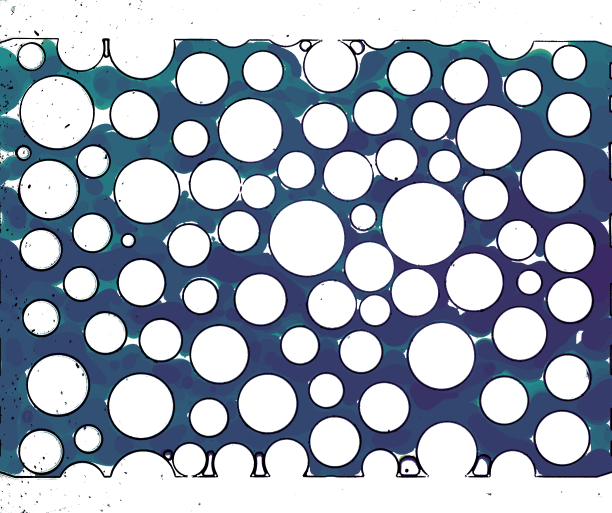}%
  }%
  \\%
  \subfloat[\label{fig:time-ensemble-5-10} $\ensembleParams{-5}{10}$]{%
    \includegraphics[width=\ensembleSize\columnwidth,trim={0cm 1.2cm 0cm 0.6cm},clip]{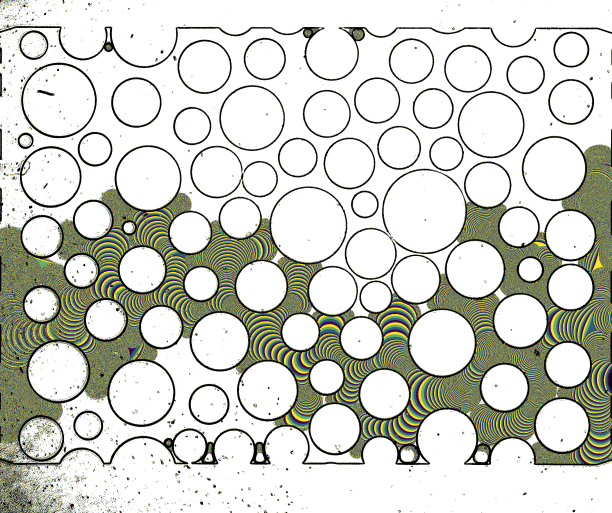}%
  }%
  \hspace{1em}%
  \subfloat[\label{fig:time-ensemble-4-10} $\ensembleParams{-4}{10}$]{%
    \includegraphics[width=\ensembleSize\columnwidth,trim={0cm 0.8cm 0cm 1cm},clip]{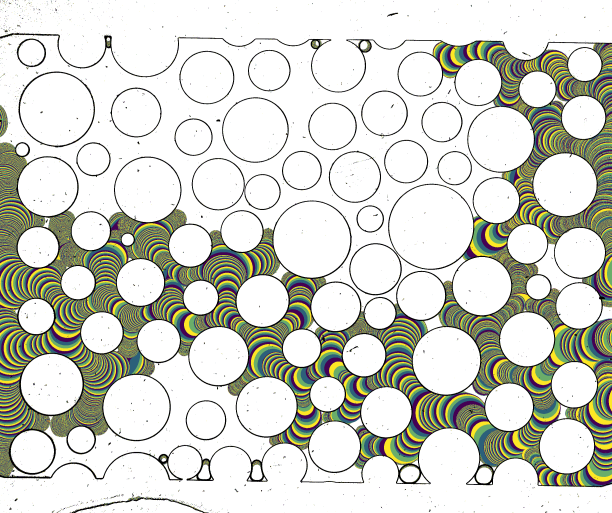}%
  }%
  \hspace{1em}%
  \subfloat[\label{fig:time-ensemble-3-10} $\ensembleParams{-3}{10}$]{%
    \includegraphics[width=\ensembleSize\columnwidth,trim={0cm 1.55cm 0cm 0.25cm},clip]{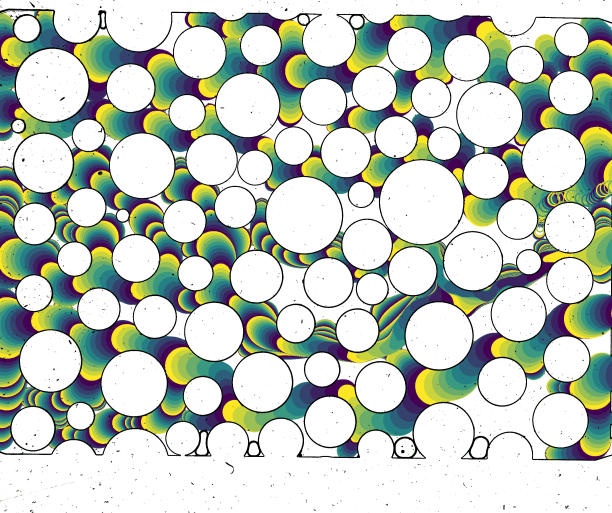}%
  }%
  \hspace{1em}%
  \subfloat[\label{fig:time-ensemble-2-10} $\ensembleParams{-2}{10}$]{%
    \includegraphics[width=\ensembleSize\columnwidth,trim={0cm 0.95cm 0cm 0.85cm},clip]{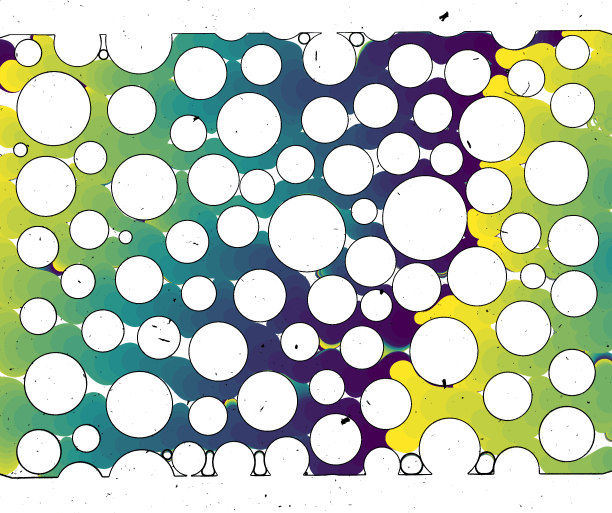}%
  }%
  \caption{%
    Time map for the ensemble dataset with varying capillary number and viscosity ratio.
    The discrete time frame, i.e., the index of the input image, is periodically mapped to color~(\viridis).
    The resulting frequency is normalized by varying the number of colors in the categorical color map depending on the frequency of the camera.
    This means that one period always relates to the same physical time, i.e., $2s$ per period.
  }%
  \label{fig:time-ensemble}%
\end{figure*}

\begin{figure}[t]%
  \centering%
  \subfloat[\label{fig:time-circular}]{%
    \includegraphics[width=0.49\columnwidth,trim={0cm 2cm 0cm 1.5cm},clip]{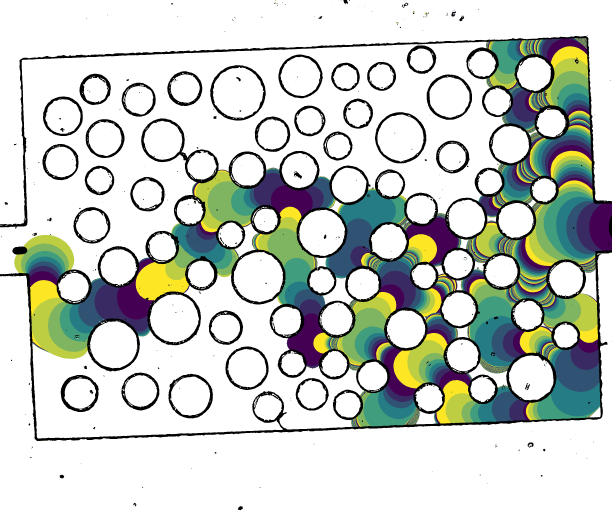}%
  }%
  \hfill%
  \subfloat[\label{fig:time-octagonal}]{%
    \begin{tikzpicture}%
      \node[anchor=south west, inner sep=0] (image) at (0,0) {\includegraphics[width=0.49\columnwidth,trim={0cm 2cm 0cm 1.5cm},clip]{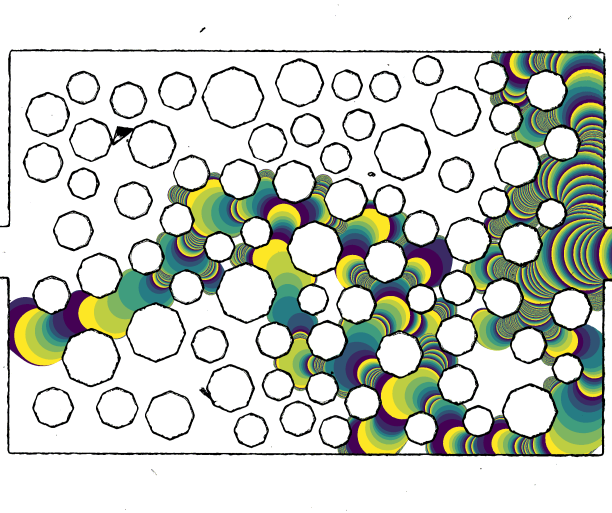}};%
      \draw[gred,ultra thick] (3.05,0.1) rectangle (4.05,0.9);%
    \end{tikzpicture}%
  }%
  \\%
  \subfloat[\label{fig:graph-octagonal-jumps}]{%
    \includegraphics[width=0.49\columnwidth,trim={0cm 1cm 0cm 1cm},clip]{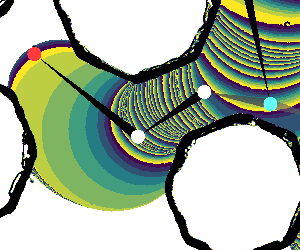}%
  }%
  \hfill%
  \subfloat[\label{fig:time-octagonal-normalized}]{%
    \includegraphics[width=0.49\columnwidth,trim={0cm 2cm 0cm 1.5cm},clip]{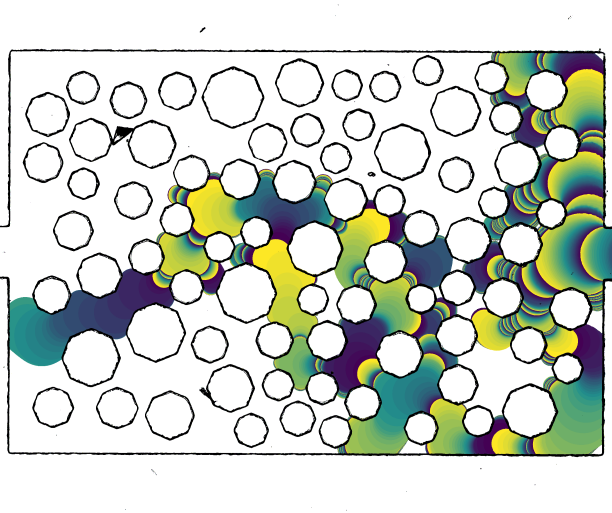}%
  }%
  \caption{%
    Time map for the ensemble dataset with varying solids: \subref{fig:time-circular}~circular, and \subref{fig:time-octagonal}~octagonal solid structure.
    The discrete time frame is periodically mapped to color~(\viridis).
    Its frequency indicates local velocity but depends on the frame rate of the capturing camera.
    Difference in saturation between \subref{fig:time-circular}~circular, and \subref{fig:time-octagonal}~octagonal solid structure is marked in red.
    \subref{fig:time-octagonal-normalized}~Adjusted color map for the octagonal dataset to match the frequency of the circle dataset.
    One period of~\subref{fig:time-circular}~and~\subref{fig:time-octagonal-normalized} now covers the same physical time.
    \subref{fig:graph-octagonal-jumps}~Zoom-in on the displacement graph with preserved nodes~(white) at velocity jumps for the octagonal dataset.
  }%
  \label{fig:time-geom}%
\end{figure}

\textbf{Time Map.}
Although the time map is only a preprocessing step towards the abstract graph visualization, it already proves useful for identifying interesting flow behavior.
Within a single image, one can easily observe different velocity distributions, and by this identify regions of small or large velocities.
Therefore, we can already differentiate between experiments with very regular velocity distributions~(e.g.,~\autoref{fig:time-ensemble-2-10}) and datasets containing large velocity ``jumps''~(e.g.,~\autoref{fig:time-octagonal-normalized}).
By additionally normalizing time, i.e., adjusting the coloring of the time map such that $1s$ in physical time always refers to the same color gradient, we also gain comparability between different experiments that have been captured with different frame rates.
Through interactive manipulation of the selected time frame and corresponding coloring, the user can also animate the progressing flow in our time map.

\begin{figure}[t]%
  \centering%
  \subfloat[\label{fig:graph-circular-simplified}]{%
    \vspace{-1em}%
    \includegraphics[width=0.49\columnwidth,trim={0cm 8cm 0cm 8cm},clip]{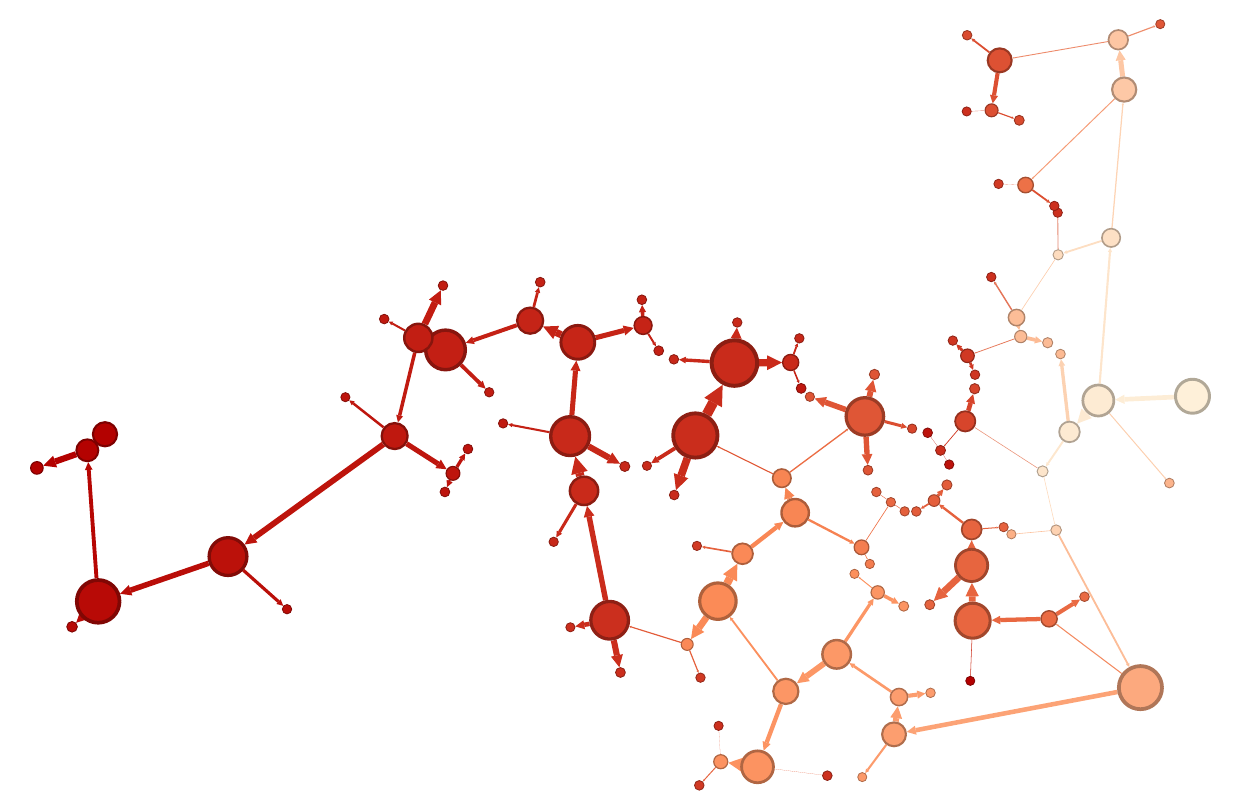}%
  }%
  \hfill%
  \subfloat[\label{fig:graph-octagonal-simplified}]{%
    \vspace{-1em}%
    \includegraphics[width=0.49\columnwidth,trim={0cm 8cm 0cm 8cm},clip]{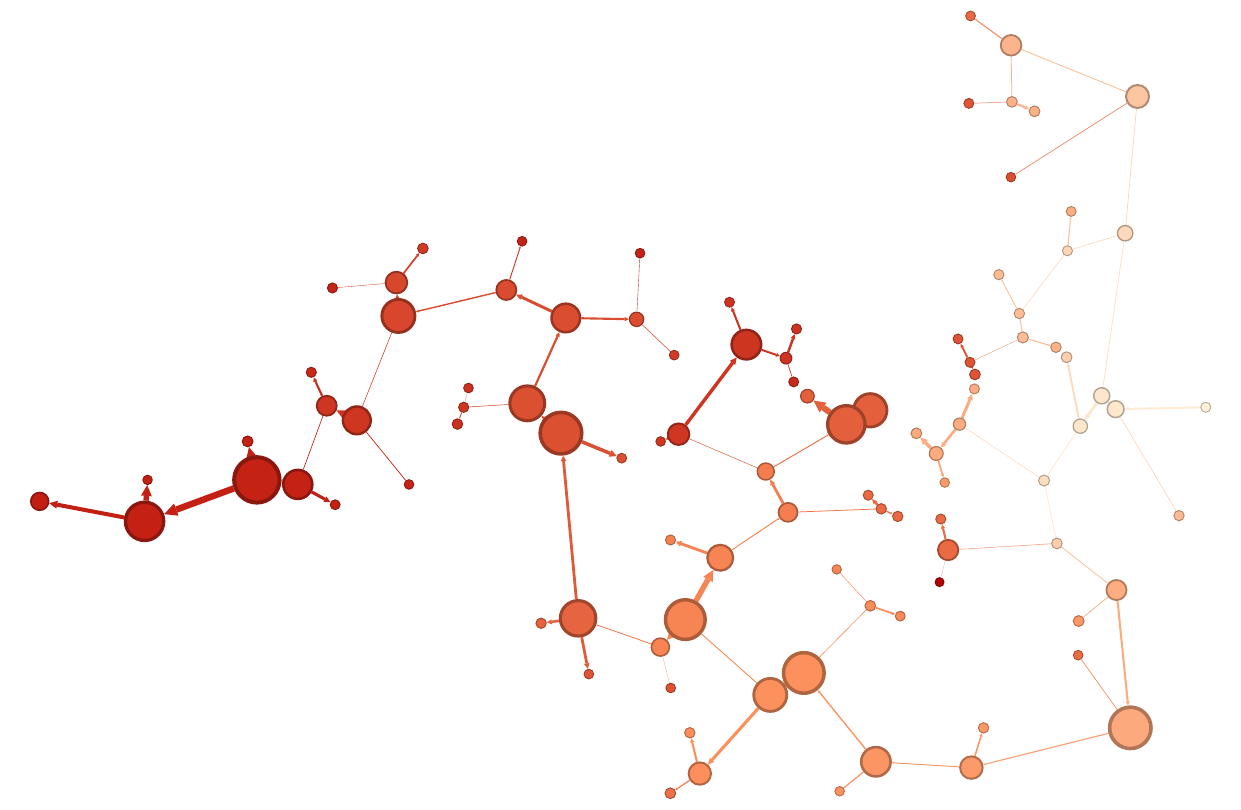}%
  }%
  \\%
  \subfloat[\label{fig:graph-circular-mainchannel}]{%
    \vspace{-1em}%
    \includegraphics[width=\columnwidth,trim={0cm 12cm 0cm 12cm},clip]{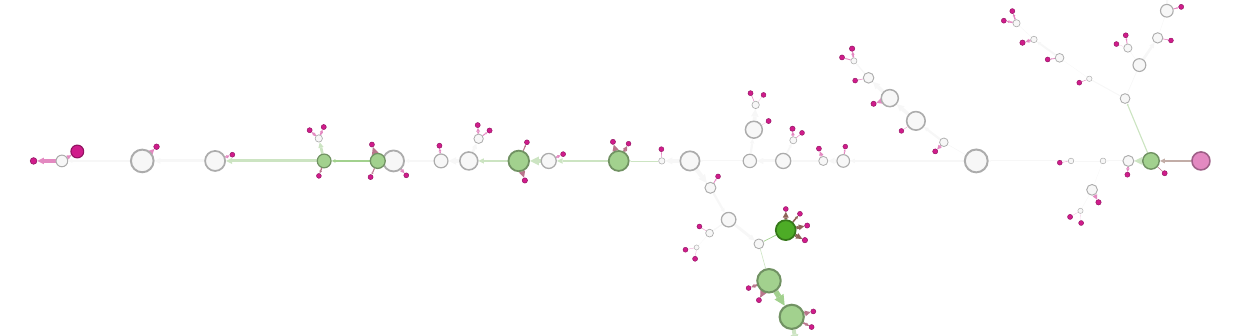}%
  }%
  \\%
  \subfloat[\label{fig:graph-octagonal-mainchannel}]{%
    \vspace{-1em}%
    \includegraphics[width=\columnwidth,trim={0cm 13cm 0cm 13cm},clip]{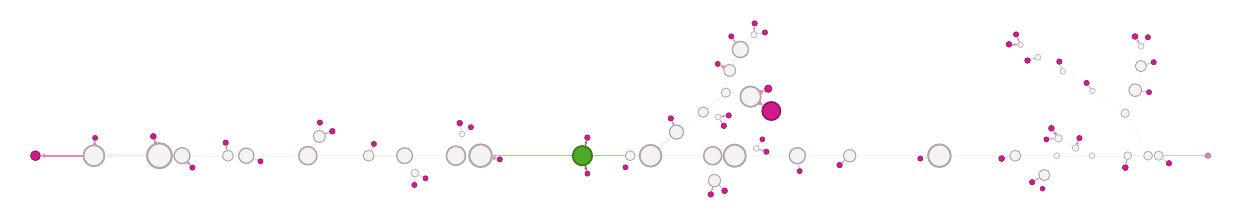}%
  }%
  \caption{%
    Displacement graph for the \subref{fig:graph-circular-simplified}~circular, and \subref{fig:graph-octagonal-simplified}~octagonal solid structure.
    The node positions coincide with the centers of mass of the corresponding flow fronts, and flow front area is mapped to node size.
    Further, velocity is depicted by the width of the edges, and time is mapped to color~(\reds) for nodes and edges.
    Breakthrough graph for the \subref{fig:graph-circular-mainchannel}~circular, and \subref{fig:graph-octagonal-mainchannel}~octagonal solid structure.
    It shows the same displacement graph with different layout, where nodes of the main channel from inlet to outlet at breakthrough are located on a horizontal line.
    Along this main channel, the distances between the nodes are preserved.
    Angles, distances and positions for all other nodes are changed from using a force-based layouting algorithm.
    Color is now mapped to the degree of outgoing edges~(\categories) to highlight differences in branching of the individual fingers.
  }%
  \label{fig:graph-layout}%
\end{figure}

\textbf{Displacement and Breakthrough Graph.}
The displacement graph provides an abstraction of the flow displacement process.
Here, we can map arbitrary quantities, such as the metrics introduced in~\autoref{sec:flow-segmentation}, to visual channels of the graph.
By mapping time to node color, flow front area to node size and velocity to edge width, and by keeping the graph nodes at their original positions, we already get a good impression about the distribution of fluid flow~(\autoref{fig:graph-circular-simplified}~and~\ref{fig:graph-octagonal-simplified}).
Some aspects, such as how far the fluid progresses into throats of the porous medium, are easily comparable in the graphs.
For example, it is observable that for the circular dataset~(\autoref{fig:graph-circular-simplified}), the fluid does not progress as far into throats as for the octagonal dataset~(\autoref{fig:graph-octagonal-simplified}).
Applying a different layout to the graph, here, by keeping the main channel on a horizontal line, we can get a better impression of branching when comparing different graphs.
As such, the breakthrough graph for the circular solid geometry~(\autoref{fig:graph-circular-mainchannel}) has an additional large branch from the main channel, as compared with the octagonal dataset~(\autoref{fig:graph-octagonal-mainchannel}).
Further, from coloring by outgoing nodes, we can see that in the circular dataset there are nodes with even up to four sinks.
In the triangular dataset~(\autoref{fig:graph-triangular-mainchannel}), we can additionally more easily identify loops and disconnected parts of the graph, compared to original layout~(\autoref{fig:graph-triangular-simplified}).

\begin{figure}[t]%
  \centering%
  \subfloat[\label{fig:framework-merges-overview}]{%
    \begin{tikzpicture}
      \node[anchor=south west, inner sep=0] (image) at (0,0) {\includegraphics[width=0.6\columnwidth,trim={3cm 1cm 4.3cm 0cm},clip]{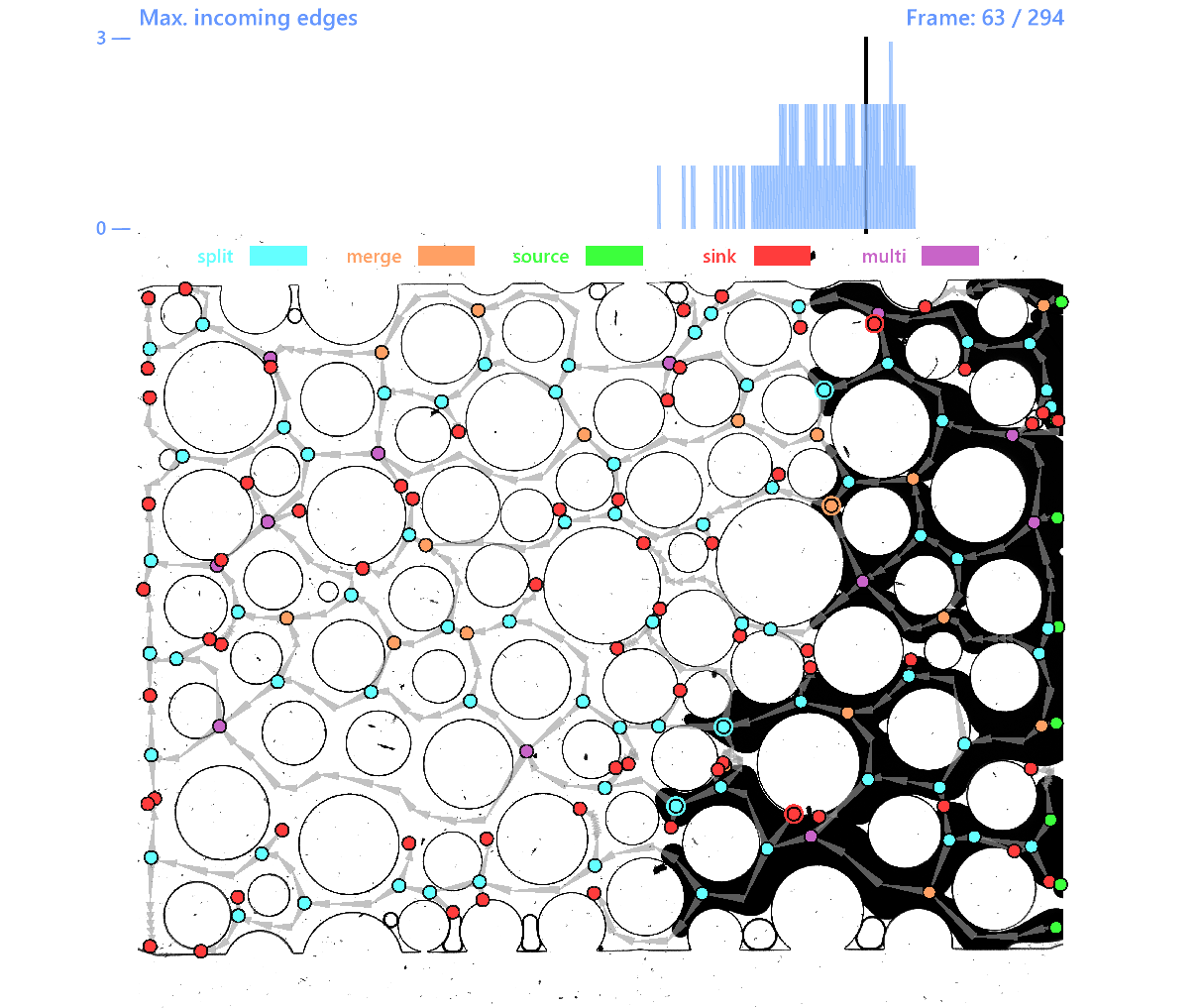}};%
      \draw[gred,very thick] (3.54,2.21) rectangle (4.31,2.98);%
      \draw[gblue,very thick] (1.97,0.85) rectangle (2.74,1.61);%
    \end{tikzpicture}%
  }%
  \hfill%
  \subfloat[\label{fig:framework-merges-steps}]{%
    \begin{minipage}{0.177\columnwidth}%
      \centering%
      \includegraphics[width=\columnwidth]{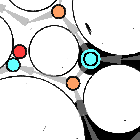}%
      \\[0.5em]%
      {\setlength{\fboxsep}{0pt}\setlength{\fboxrule}{1pt}\color{gred}\fbox{\includegraphics[width=\columnwidth]{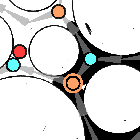}}}%
      \\[0.5em]%
      \includegraphics[width=\columnwidth]{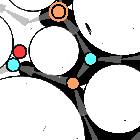}%
    \end{minipage}%
  }%
  \hspace{0.5em}%
  \begin{tikzpicture}%
    \draw[anchor=south west,inner sep=0,black,thick,<-] (0,0) -- (0,5.2);%
  \end{tikzpicture}%
  \hfill%
  \subfloat[\label{fig:framework-retracting-steps}]{%
    \begin{minipage}{0.177\columnwidth}%
      \centering%
      \includegraphics[width=\columnwidth]{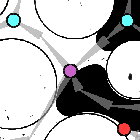}%
      \\[0.5em]%
      {\setlength{\fboxsep}{0pt}\setlength{\fboxrule}{1pt}\color{gblue}\fbox{\includegraphics[width=\columnwidth]{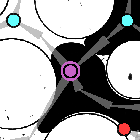}}}%
      \\[0.5em]%
      \includegraphics[width=\columnwidth]{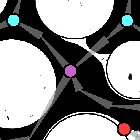}%
    \end{minipage}%
  }%
  \caption{%
    Interactive tool showing the ensemble dataset with $\ensembleParams{-2}{10}$, where there exists a node with two incoming edges.
    \subref{fig:framework-merges-overview}~Overview with highlighted regions where a merge event is expected in two different time steps.
    \subref{fig:framework-merges-steps}~Steps right before, at, and right after the merger~(orange node).
    \subref{fig:framework-retracting-steps}~Differently, a merger seems to occur but the fluid arriving from the top right then retracts again~(purple node).
  }%
  \label{fig:framework-merges}%
\end{figure}

\textbf{Interactive Framework.}
Our interactive visual analysis tool combines the advantages of the abstraction of the graph with the computed metrics.
Especially, we can use our approach to investigate outliers, such as jumps in velocities.
For this, we can show the respective metric as a bar chart, and by selecting an individual frame can visualize the corresponding graph and original input image.
Another example is that the domain scientists are interested in knowing if fingers merge in the process of the displacement process.
Therefore, we looked at different time steps of the $\ensembleParams{-2}{10}$ dataset and found two examples of merging represented in the graph structure by visualizing as metric the maximum number of incoming edges per frame.
The first one simply shows two fingers merging into one single finger~(\autoref{fig:framework-merges-overview}~and~\ref{fig:framework-merges-steps}).
However, the second proved to be rather unexpected.
Here, it seems that two fingers are merging~(\ref{fig:framework-retracting-steps}), but looking at later time steps, the finger coming from the top right retracts.
According to our domain scientists, this is probably because the merger never happened due to a thin film of defending fluid separating the fingers.
This is not a limitation of our approach but of the resolution of the camera.
However, this seems to be a very interesting occurrence and was only identified during interactive visual analysis.

\section{Domain Insights}
\label{sec:results-domain}

\revision{Our work has been developed in an iterative manner together with domain scientists.
Although no specific tasks were defined, they identified interesting metrics and phenomena, which we incorporated into our visualizations.
For the goal of gaining new insights from the datasets, we provided them with iteratively improved and extended prototypes.
Because they were involved closely in this process with incremental updates, they did not encounter a steep learning curve.}

\textbf{Flow and transport in porous media.}
The definition and effective way of defining and using the capillary number as a global metric has been a matter of debate.
With the visualization potential proposed in this work, we can clearly show that the capillary number is indeed a metric which is ill-defined.
Depending on the desired global capillary number, and in combination with the viscosity of the invading phase, the boundary flux conditions were accordingly adjusted.
However, depending on the flow conditions, meaning if flow is capillarity- or viscosity-dominated, the sweeping efficiency of the invading phase during primary drainage varies.
This has been described by Lenormand et al.~\cite{Lenormand1988} and his classification of flow in the capillary number-viscosity ratio space.
In~\autoref{fig:time-ensemble}, this is shown by the value of the breakthrough saturation, and the distribution of phases at breakthrough.
This has already an effect on the effective capillary number, since one of the assumptions for the definition of the capillary number is that the whole cross-section of the porous medium will be invaded.
The inaccuracy of this assumption can easily be seen in~\autoref{fig:time-ensemble}.
Under capillarity-driven flow conditions, as in~\autoref{fig:time-ensemble-5-0.2}, \ref{fig:time-ensemble-5-1} and~\ref{fig:time-ensemble-5-10}, meaning small capillary numbers, the breakthrough saturation is expected to be systematically smaller than this for a viscosity-driven flow regime, as in~\autoref{fig:time-ensemble-3-0.2}, \ref{fig:time-ensemble-3-1} and~\ref{fig:time-ensemble-3-10}.
That, in practice, means that a smaller percentage of the available pore space will participate in flow.
Even though the assumption of a capillarity-driven regime is in general applicable, invasion effects are not necessarily equally slow.
In~\autoref{fig:time-ensemble-5-0.2}, \ref{fig:time-ensemble-5-1} and \ref{fig:time-ensemble-5-10}~this can be seen in detail, where the same capillary number for different viscosity ratios induces different pore scale velocities.
Depending on the geometrical local constraints, it can happen that the local capillary number is very different to the one globally assumed.
This can locally change the behavior of the system from capillarity- to viscosity-driven flow, making any assumption for the global system behavior irrelevant.

In~\autoref{fig:time-circular}, the network used in the work of Yiotis et al.~\cite{Yiotis2021} is shown, where the grains have a cylindrical 3D shape.
In~\autoref{fig:time-octagonal}, the exact same grain topology is shown, with the only difference being that the grains now have an octagonal shape, inscribed in the circles (in 2D) of~\autoref{fig:time-circular}.
Both experiments took place under the same boundary capillary number conditions, taking under consideration a depth difference of ${10 \mu m}$.
At a first glance, the distributions of the invading phase for both pore geometries seem identical.
However, a more detailed look with the use of the developed visualization tool reveals a number of interesting and useful differences.
For instance, under analogous boundary flux conditions, so as to ensure a capillarity-dominated regime, the breakthrough saturation of the circular grains is higher than the one for the octagonal ones.
The main difference is marked in red in~\autoref{fig:time-octagonal}.
This can be arbitrarily assigned to random roughness of the two microfluidic chips.
However, a closer look in~\autoref{fig:time-octagonal-normalized} reveals that most of the interfaces in the octagonal grains get pinned at the corners of the grains.
The corresponding interfaces get "stuck" at the locations where the contact angle can potentially become discontinuous due to the change in the angle of the grain.
Any further increase of the local pressure initiates violent transitions, known as Haines jumps, from one time step to the other, which is also depicted with the sequential invasion steps in the corresponding images.
With this visualization approach, we can easily map the sequential steps of flooding, and how the grain shape affects the evolution of flow both in terms of local capillary numbers, which in the case of the octagonal grains is locally higher, and breakthrough saturation.

Each displacement event is critically affected by the boundary conditions, like the applied flux or pressure gradient, the distribution of grains as geometrical entities, the distribution of sizes for grains/pores, and the coordination number, meaning the number of connections for a given pore body, among others.
Especially when such a displacement event is simultaneously carrying solutes, like viruses, pathogens, surfactants, and other advective driven particles, the distribution of the carrying phase with respect to the direction of flow is of crucial importance.
As shown in various works in the literature~\cite{Chen2021,Hasan2020}, the distribution of the carrying phase can significantly affect advection-dispersion via the formation of stagnant zones, meaning zones that can be "fed" with solutes under only diffusion.
Such regions are most commonly shaped from dead-end pores, or clusters of them.
In~\autoref{fig:graph-circular-mainchannel}, \ref{fig:graph-octagonal-mainchannel}, and~\ref{fig:graph-triangular-mainchannel}, the degree of branching with respect to the direction of flow is shown for the circular, octagonal and triangulation pore geometries, respectively.
With the use of such a visualization approach, it is made possible to get an estimation of the dispersion of the saturation of the invading phase saturation.
In the case that the same phase is the one carrying the solutes, it will also provide an estimation in a qualitative way regarding the formation and existence of stagnant zones, due to the existence of dead-end pores, meaning pores not connected to the outflow.
\revision{For this, the main channel graph provides complementary information, as the deviation of the distribution of the invading phase from the direction of flow.
Branching of the invading phase in a different direction than of the displacement creates dead-end pores, which will eventually favor the creation of stagnant zones.
For this process it is of higher importance to show that this branching exists and how different pore geometries can favor or not the stagnation of dispersed solutes or pathogens than the displacement time scale itself.}

From~\autoref{fig:metrics-circular-area}, \ref{fig:metrics-octagonal-area}, \ref{fig:metrics-triangular-area}, flow front area versus time is plotted for the circular, octagonal and triangulation pore space, respectively.
It can be clearly seen that there are times in the case of the octagonal network where nothing is actually happening, since the interfaces are stuck in the corners of the grains.
Contrarily, in the other two networks, there is always a production of area occupied by the invading phase, since there is a continuous movement of the corresponding phase in the pore space.
This is also depicted in the second row~(\autoref{fig:metrics-circular-velocity}, \ref{fig:metrics-octagonal-velocity}, \ref{fig:metrics-triangular-velocity}), where the velocity is again continuous in the case of the circular and triangulation network, but not for the octagonal one.
Accordingly, the flow front produced from the octagonal network is systematically less than this for the cylindrical network, since the produced interfaces tend to jump from one state to the other before expanding too much so as to excessively increase their volume.
This general image remains consistent also in the rest of the rows, where fluid-fluid~(\autoref{fig:metrics-circular-fluid}, \ref{fig:metrics-octagonal-fluid}, \ref{fig:metrics-triangular-fluid}) and fluid-solid~(\autoref{fig:metrics-circular-solid}, \ref{fig:metrics-octagonal-solid}, \ref{fig:metrics-triangular-solid}) interfacial area is plotted, as well as the number of fingers.
Especially the number of fingers between the two nearly identical pore spaces is of special interest~(\autoref{fig:metrics-circular-fingers}, \ref{fig:metrics-octagonal-fingers}, \ref{fig:metrics-triangular-fingers}).
The discontinuity in the contact angle in the case of the octagonal grains constrains both the number of the fingers, since there is enough time for the emerging fingers to compete until one of them is released, against the cylindrical grains case where this constrain is not present anymore, and fingers are allowed to move more freely.

\begin{figure}[t]%
  \centering%
  \def\metricsWidth{0.28}%
  \newcommand{\labelInside}[2]{
    \begin{tikzpicture}%
      \node[anchor=south west] (image) at (0,0) {\includegraphics[width=\metricsWidth\columnwidth,trim={0.8cm 1.8cm 0cm 0cm},clip]{#1}};%
      \node at (0.45,0.8) {\subfloat[\label{#2}]{}};%
    \end{tikzpicture}%
  }%
  \begin{tikzpicture}%
    \node[anchor=south west] at (0,0) {~~~};%
    \node[anchor=west,rotate=90] at (0,0.45) {\tiny{area}};%
  \end{tikzpicture}%
  \hspace{2pt}%
  \labelInside{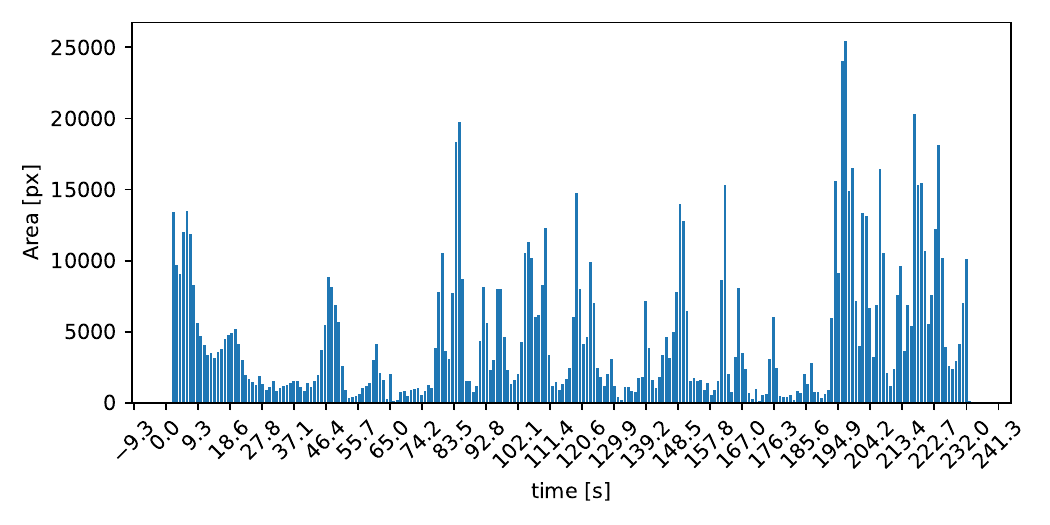}{fig:metrics-circular-area}%
  \hfill%
  \labelInside{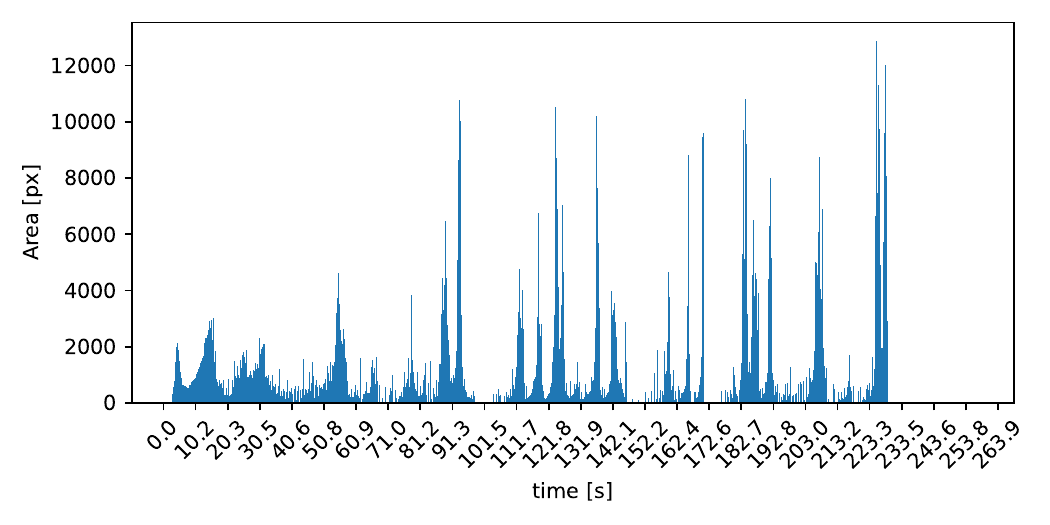}{fig:metrics-octagonal-area}%
  \hfill%
  \labelInside{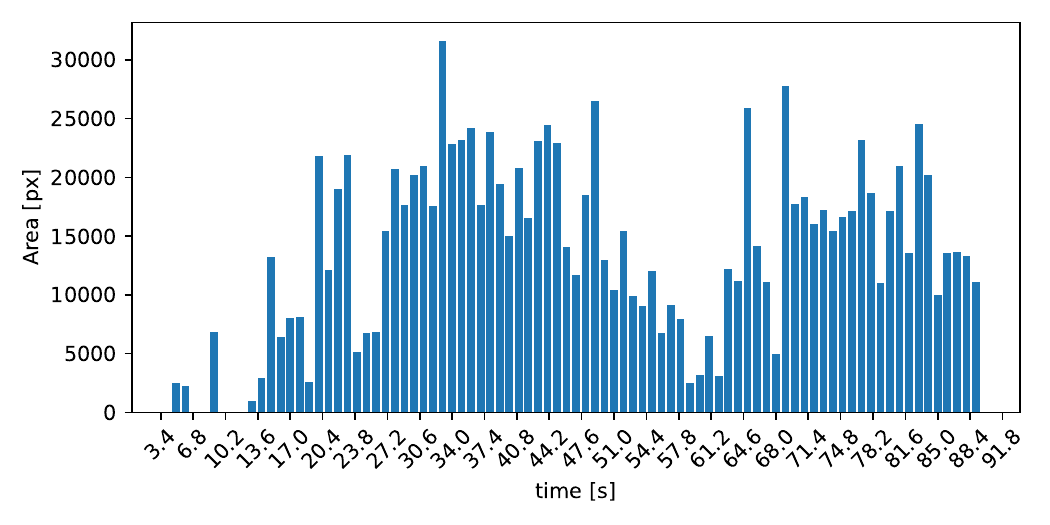}{fig:metrics-triangular-area}%
  \\[-0.5em]%
  \begin{tikzpicture}%
    \node[anchor=south west] at (0,0) {~~~};%
    \node[anchor=west,rotate=90] at (0,0.25) {\tiny{velocity}};%
  \end{tikzpicture}%
  \hspace{2pt}%
  \labelInside{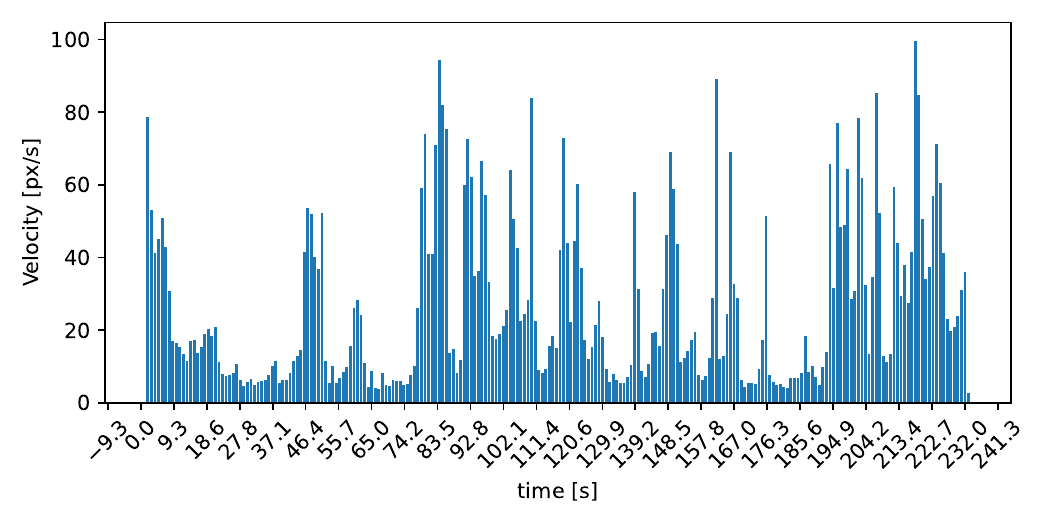}{fig:metrics-circular-velocity}%
  \hfill%
  \labelInside{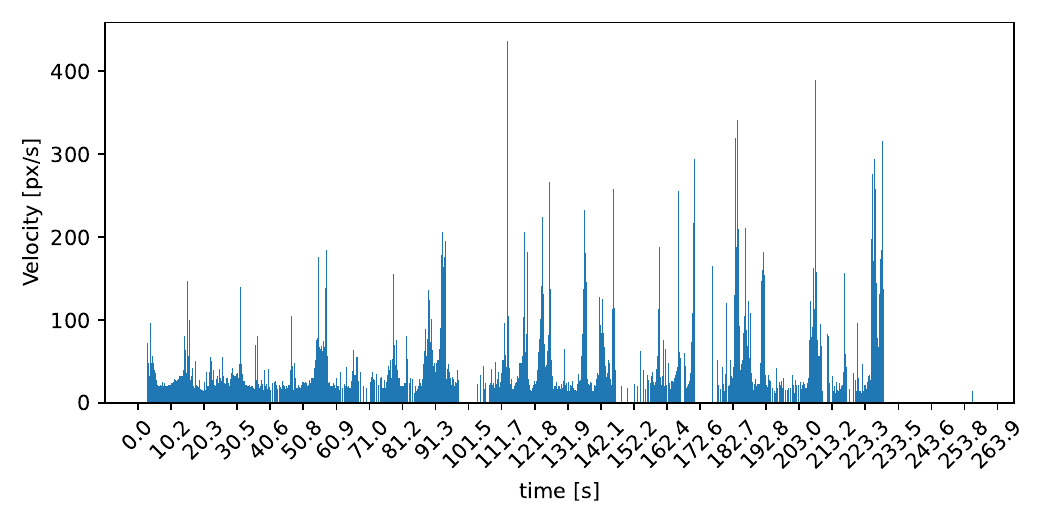}{fig:metrics-octagonal-velocity}%
  \hfill%
  \labelInside{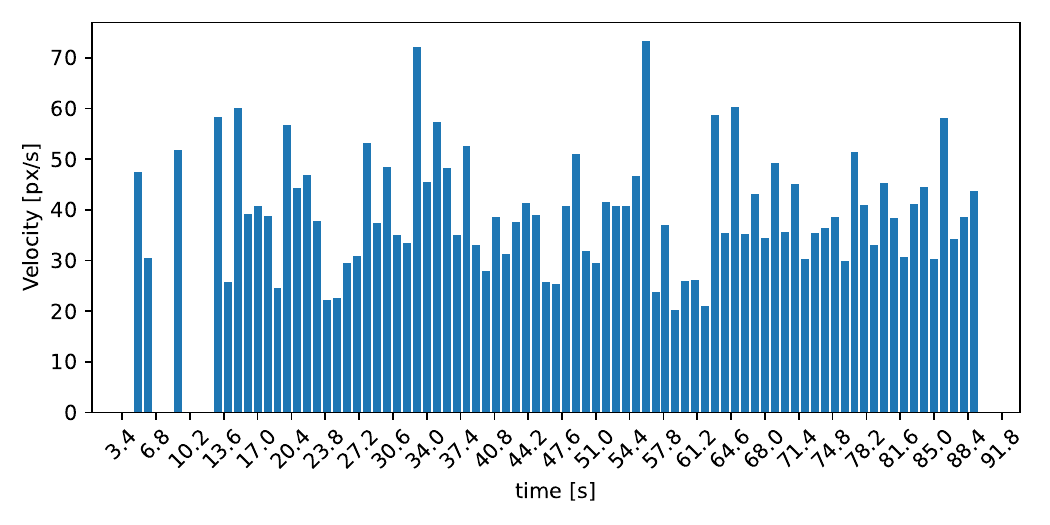}{fig:metrics-triangular-velocity}%
  \\[-0.5em]%
  \begin{tikzpicture}%
    \node[anchor=south west] at (0,0) {~~~};%
    \node[anchor=west,rotate=90] at (0,0.2) {\tiny{fluid-fluid}};%
    \node[anchor=west,rotate=90] at (0.2,0.25) {\tiny{interface}};%
  \end{tikzpicture}%
  \hspace{2pt}%
  \labelInside{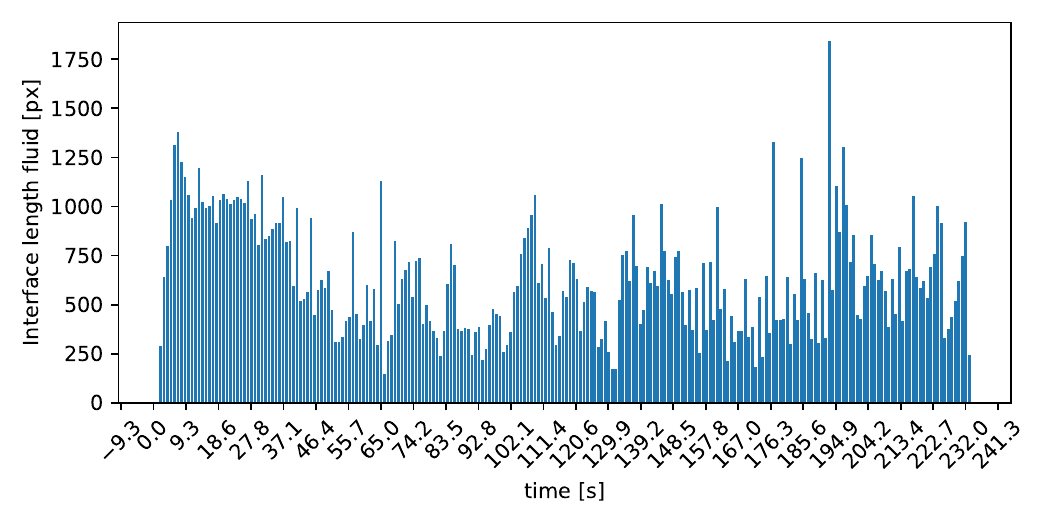}{fig:metrics-circular-fluid}%
  \hfill%
  \labelInside{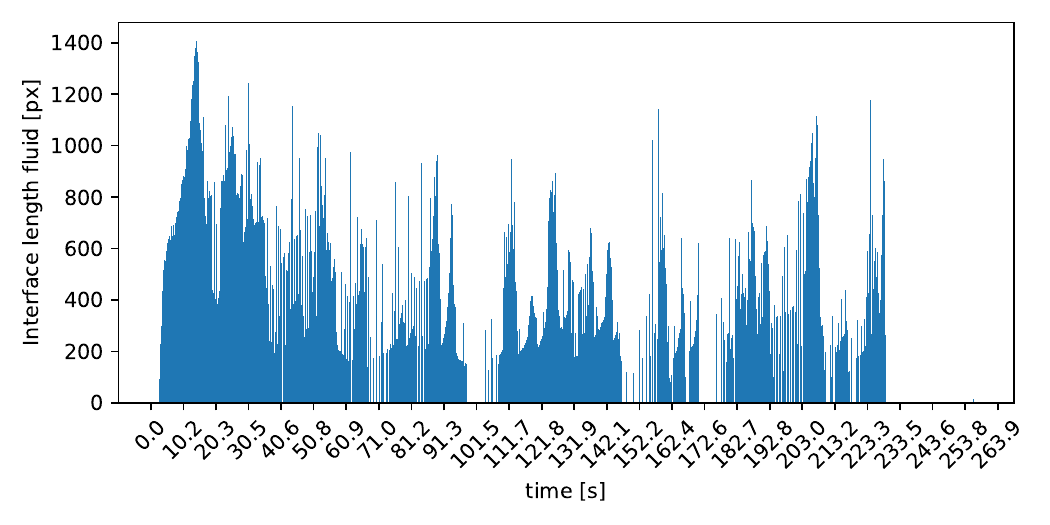}{fig:metrics-octagonal-fluid}%
  \hfill%
  \labelInside{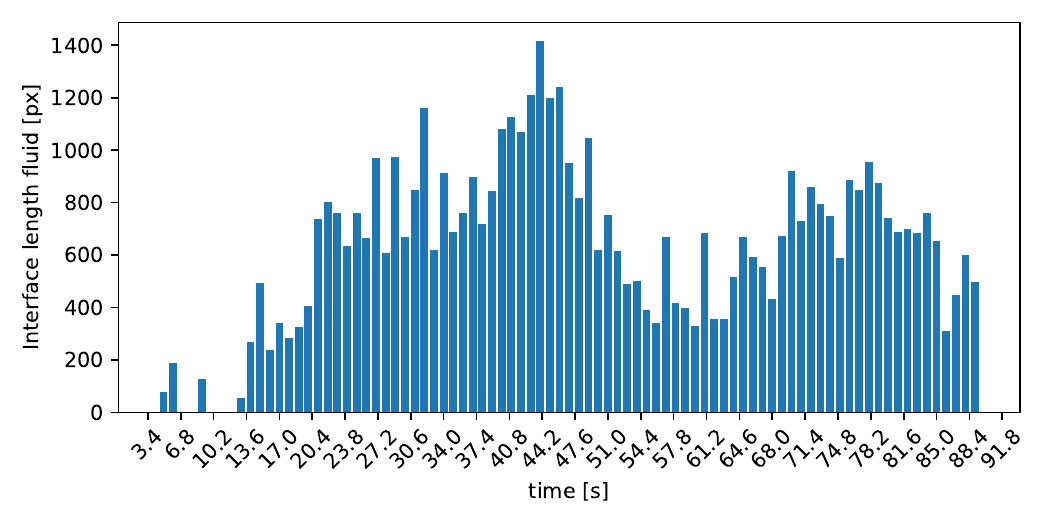}{fig:metrics-triangular-fluid}%
  \\[-0.5em]%
  \begin{tikzpicture}%
    \node[anchor=south west] at (0,0) {~~~};%
    \node[anchor=west,rotate=90] at (0,0.2) {\tiny{fluid-solid}};%
    \node[anchor=west,rotate=90] at (0.2,0.25) {\tiny{interface}};%
  \end{tikzpicture}%
  \hspace{2pt}%
  \labelInside{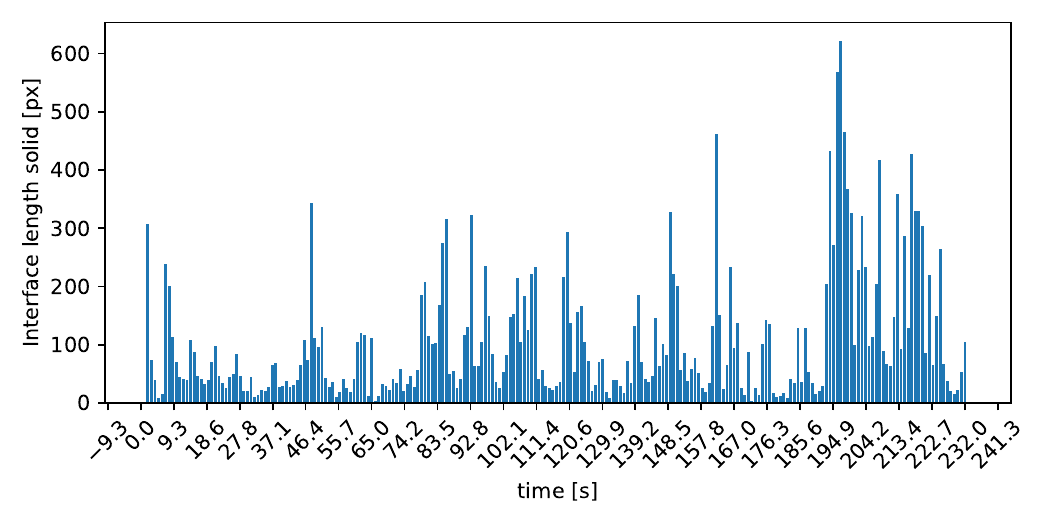}{fig:metrics-circular-solid}%
  \hfill%
  \labelInside{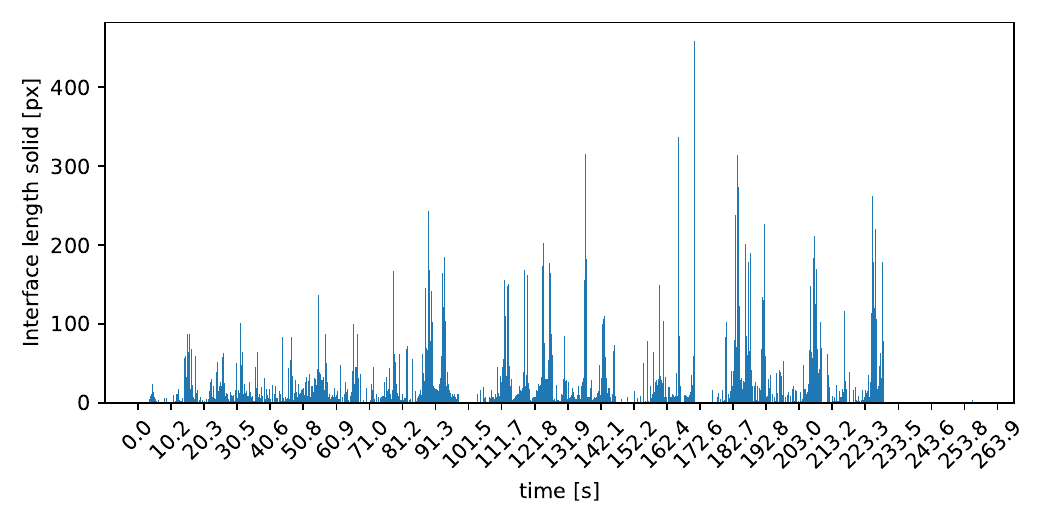}{fig:metrics-octagonal-solid}%
  \hfill%
  \labelInside{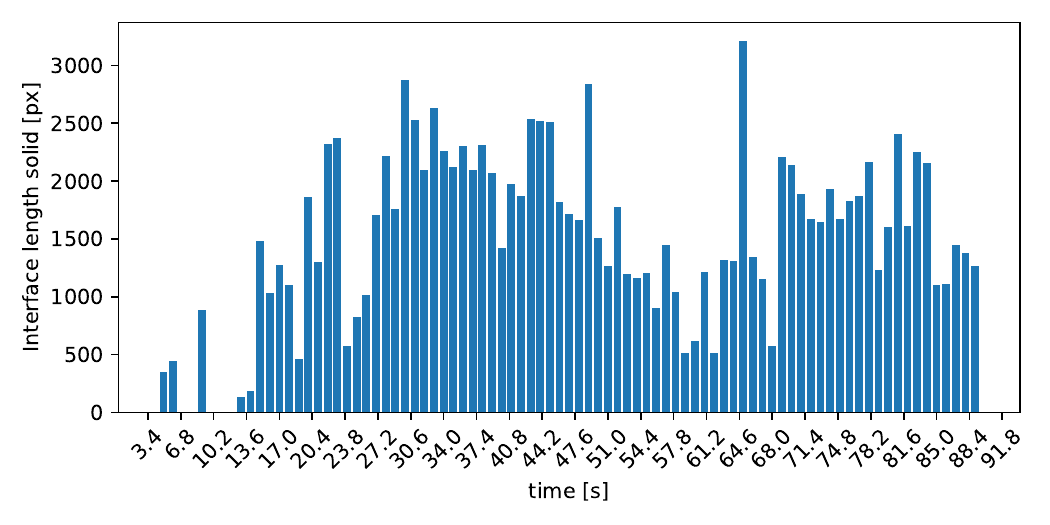}{fig:metrics-triangular-solid}%
  \\[-0.5em]%
  \begin{tikzpicture}%
    \node[anchor=south west] at (0,0) {~~~};%
    \node[anchor=west,rotate=90] at (0,0.3) {\tiny{fingers}};%
  \end{tikzpicture}%
  \hspace{2pt}%
  \labelInside{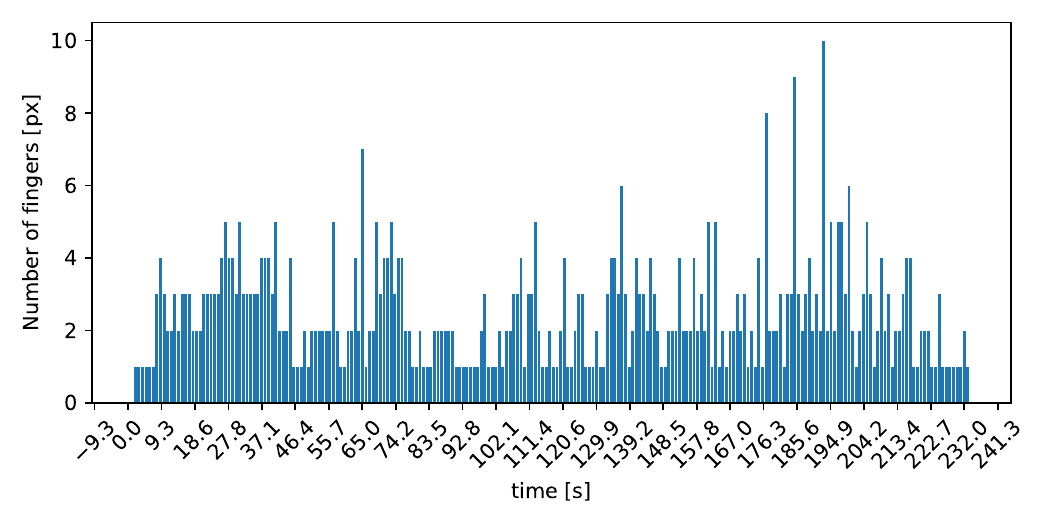}{fig:metrics-circular-fingers}%
  \hfill%
  \labelInside{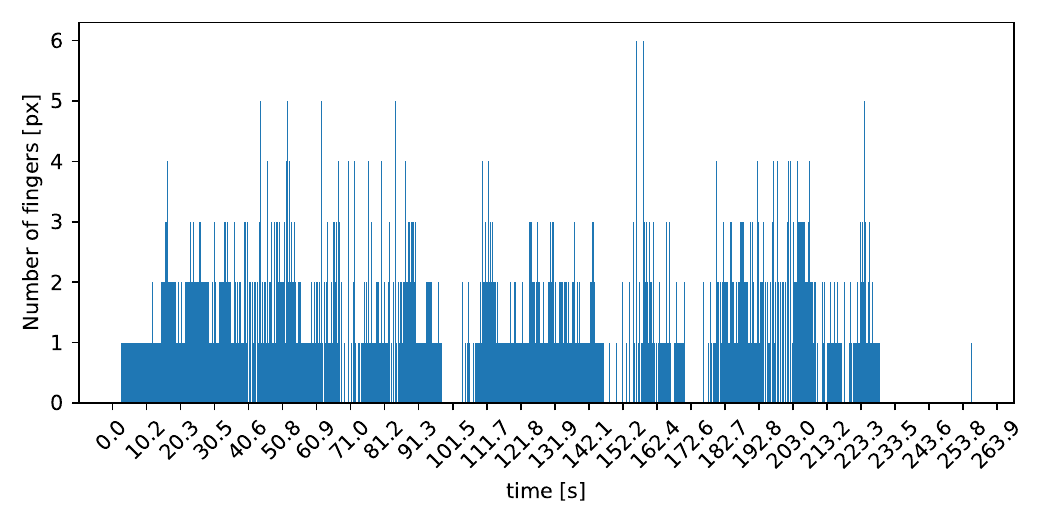}{fig:metrics-octagonal-fingers}%
  \hfill%
  \labelInside{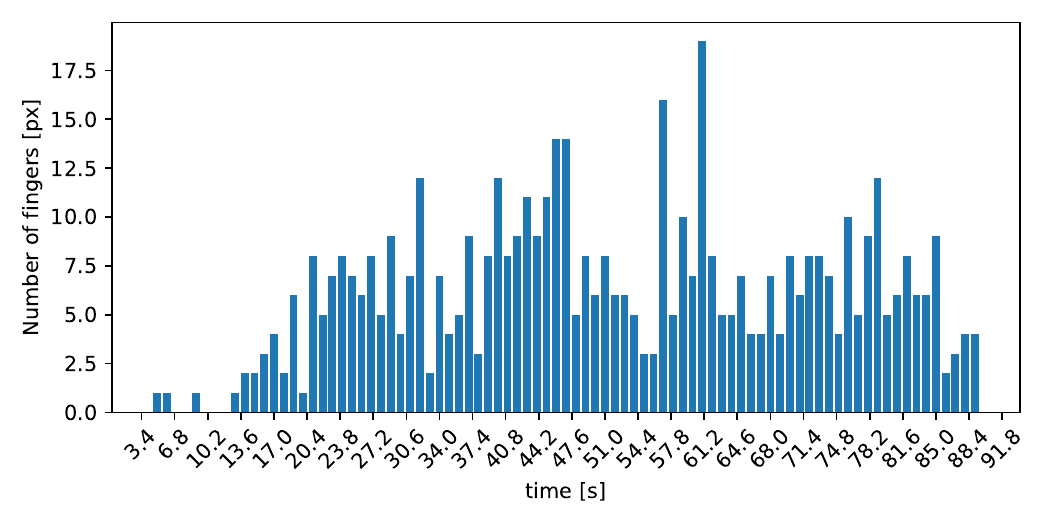}{fig:metrics-triangular-fingers}%
  \vspace{4pt}
  \caption{%
    Metrics for the different geometries: circular~(left), octagonal~(center), and triangular~(right).
    \subref{fig:metrics-circular-area}--\subref{fig:metrics-triangular-area}~Area flooded by the invading fluid in pixel.
    \subref{fig:metrics-circular-velocity}--\subref{fig:metrics-triangular-velocity}~Approximated velocity from the Hausdorff distance in pixel per second.
    \subref{fig:metrics-circular-fluid}--\subref{fig:metrics-triangular-fluid},~\subref{fig:metrics-circular-solid}--\subref{fig:metrics-triangular-solid}~Number of pixel of the fluid-fluid or fluid-solid interface at the corresponding time step.
    \subref{fig:metrics-circular-fingers}--\subref{fig:metrics-triangular-fingers}~Number of fingers, i.e., active paths in the graph.
  }%
  \label{fig:metrics}%
\end{figure}

\revision{
\textbf{Domain User Feedback.}
The developed tools are very much welcome by the domain experts, since they provide them with a fast and quantitative way to identify qualitative changes based on the boundary flux conditions, and the physical properties of the fluids involved, like the viscosity contrast.
Apart from the quantification of the flow domains in terms of time and phase distribution, the developed tools can be used to optimize the parameters of the experimental setup.
For instance, the appropriateness of a pore geometry can readily be evaluated by running a test experiment and identifying whether or not the specific pore geometry is appropriate for the specific type of experiments, whether or not dispersion is included.
Additionally, these tools also provide the domain experts with the means to identify if, for instance, the acquisition rate of the images is adequate to characterize a process, but also not too much so as to generate redundant experimental data without a significant output due to the time scale.}

\section{Discussion}
\label{sec:discussion}

At this point, we want to discuss the similarities and differences between our work and the work by Frey et al.~\cite{Frey2021}.
Although both extract graphs, they represent two different aspects of a porous medium.
Their graph is only defined by the solid structure and is limited to convex geometry.
Further, they require the application of a manually created and denoised mask.
Our graph, on the other hand, is defined directly by the flow and does not suffer from this limitation.
This means that flow \emph{within} a pore body can be covered by our graph, while theirs can only represent the pore as a whole, thus making our approach more fine-grained.
Further, node positions in our graph are less important, and thus different layouts can be used for more abstract visualization.
While both approaches visualize the domain by mapping colors to metrics, the visualization of the time map in our approach implicitly highlights velocities.
Because we do not differentiate into pores and throats, however, we are not able to visualize pore-related quantities, such as pore saturation.
Another contribution of this work is that we provide an interactive visual analysis tool.
This allows us to interactively investigate flow phenomena.
With our new approach, we can essentially gain the same information from the ensemble dataset as Frey et al.
Further, we are able to get even more insight by applying the time map and novel graph layout, analyzing, for example, branching off the main channel.

\begin{figure}[t]%
  \centering%
  \def\ensembleSize{0.33}
  \subfloat[\label{fig:graph-ensemble-4-10}]{%
    \includegraphics[width=\ensembleSize\columnwidth,trim={0cm 1cm 0cm 1cm},clip]{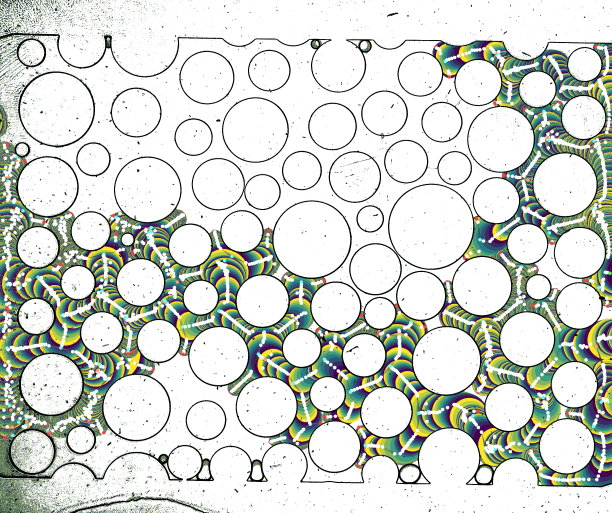}%
  }%
  \hfill%
  \subfloat[\label{fig:graph-ensemble-4-10-simplified}]{%
    \includegraphics[width=\ensembleSize\columnwidth,trim={0cm 7cm 0cm 7cm},clip]{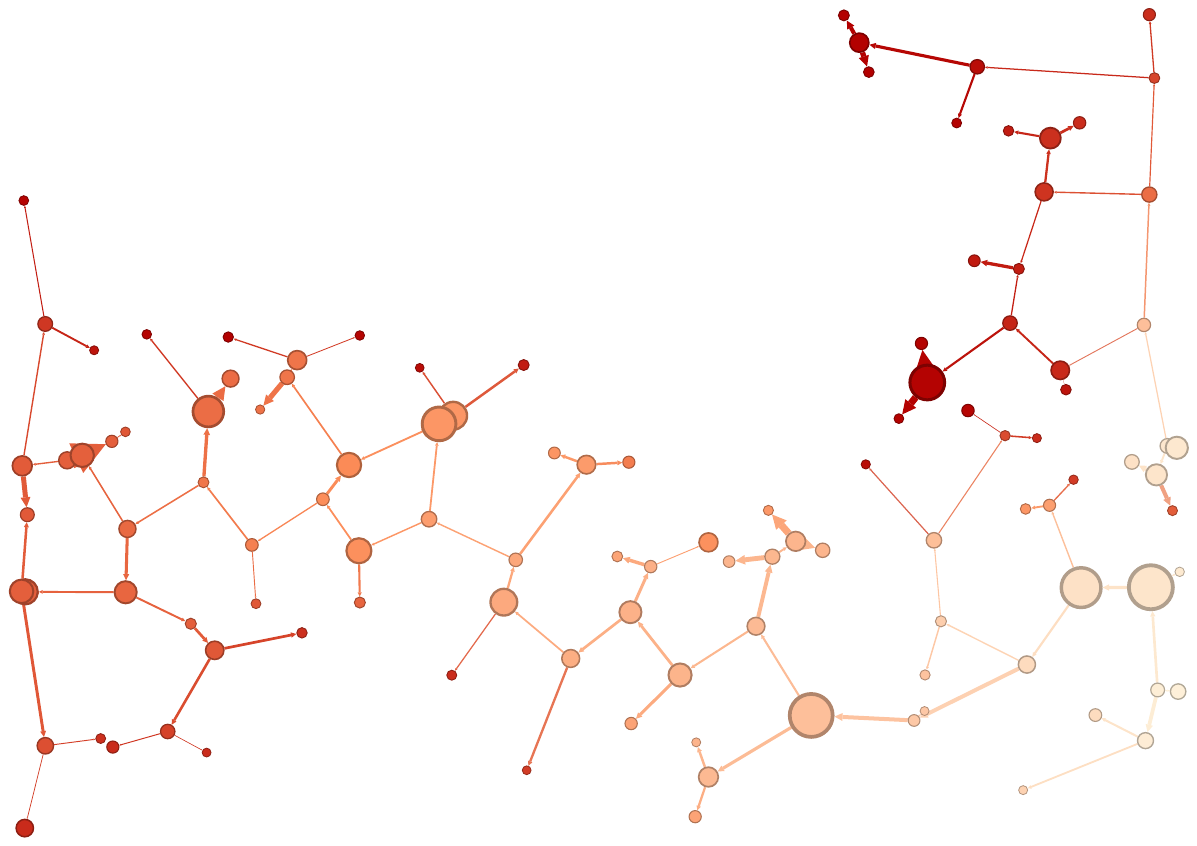}%
  }%
  \hfill%
  \subfloat[\label{fig:graph-ensemble-3-0.2}]{%
    \includegraphics[width=\ensembleSize\columnwidth,trim={0cm 1cm 0cm 1cm},clip]{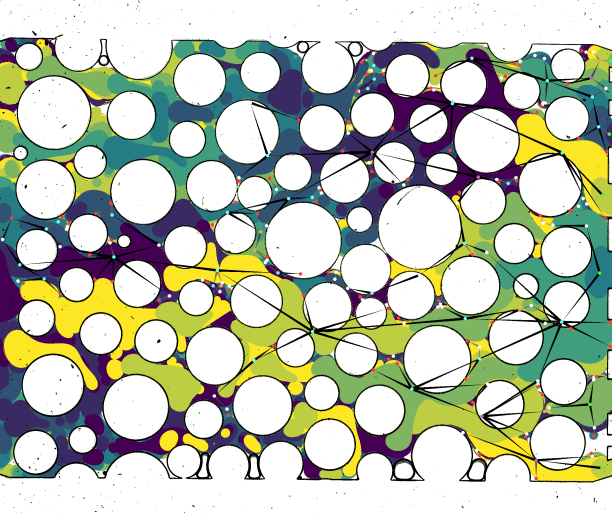}%
  }%
  \caption{%
    Two examples from the ensemble dataset with varying capillary number and viscosity ratio showing the time map and the complete displacement graph.
    \subref{fig:graph-ensemble-4-10},\subref{fig:graph-ensemble-4-10-simplified}~In the presence of noise, the resulting graph~($\ensembleParams{-4}{10}$) is still very robust, as can be seen on the left side of the domain.
    \subref{fig:graph-ensemble-3-0.2}~Our method fails for flow that is viscosity-driven~($\ensembleParams{-3}{0.2}$), as the invading fluid forms blobs and vacates previously occupied space.
  }%
  \label{fig:graph-ensemble}%
\end{figure}

As already shown in the results, our approach is applicable to a wide range of datasets of flow in porous media.
However, the main limitation is regarding multiphase phenomena, such as the formation of ``blobs''.
These cannot be captured by our time map, as the invading fluid is itself displaced.
Further, our graph abstraction would not be able to properly portray this flow behavior~(see~\autoref{fig:graph-ensemble-3-0.2}).
To handle such flow, a whole new approach is needed.
Additionally, data with too coarse temporal resolution would be difficult to handle if splits and merges occur within a single time frame, and are not detected by our graph.

\section{Conclusion}

Our goal was to create (abstract) visualizations for the analysis of ensemble datasets of porous media experiments.
To this end, we developed an approach that encodes the time of displacement as a value in a single image, called time map.
This map can directly be used for the implicit visualization of velocities by means of frequency.
Further, we introduced a method to generate displacement graphs, and---by applying a different layout---breakthrough graphs, which represent the displacement process in an abstract manner.
Finally, we provide an interactive visual analysis tool that can be used to investigate phenomena by helping relate metrics to the image data and the displacement graph. 

Applying our new visualizations, our domain scientists analyzed two ensemble datasets, one with varying capillary number and viscosity ratio, and the other with varying geometric structures.
Using the time map, they showed that a global capillary number is not an accurate descriptor, due to local capillarity changes.
Further, our visualizations helped them observe the Haines jumps in the case of the octagonal dataset, relating the behavior to the difference in geometry when comparing with circular solid structures.
Finally, they used our breakthrough graph to analyze advection-dispersion differences between ensemble members with varying geometry, thus showing the usefulness of our new approach for the analysis of porous media ensembles.

As future work we want to also consider the comparison between experiments and simulation results.
\revision{Because visual comparison is limited by graph size, we want to extend our approach by applying graph metrics.
These can be used to compare the resulting graphs quantitatively.}
Further, we want to extend our interactive graph visualization to allow the user to switch locally between simplified and original graph.
This would enable the user to get an overview from the simplified graph and then drill down to view the details from the original graph.
\revision{We would also like to extract the Reeb graph directly from our scalar time field and compare the results to our graph generation.}
Finally, another challenge will be to extend the whole approach to 3D data.

\acknowledgments{%
  The authors thank Adrian Zeyfang for the initial version of the presented approach and Gunther Weber for his valuable input on Reeb Graphs.
  Funded by the Deutsche Forschungsgemeinschaft (DFG, German Research Foundation) -- Project Number 327154368 -- SFB 1313.
  We thank the DFG for supporting this work also by funding -- EXC2075 -- 390740016 under Germany's Excellence Strategy.
  We acknowledge the support by the Stuttgart Center for Simulation Science (SimTech).
}

\bibliographystyle{abbrv-doi-hyperref}

\bibliography{paper}

\appendix
\section*{Appendix}

\begin{figure}[ht!]%
  \centering%
  \includegraphics[width=0.56\columnwidth,trim={0cm 0cm 0cm 1cm},clip]{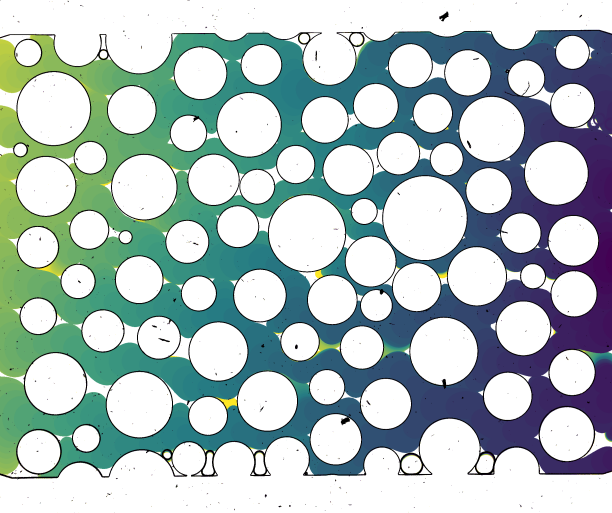}%
  \caption{%
    Time map for the same dataset as in~\autoref{fig:displacement-overview} but with a non-periodic color map.
    While we can now observe the global flow propagation, local details are lost.
    For this reason, we stick to periodic color maps (but support any color map) with the option to interactively highlight specific time steps (red areas in~\autoref{fig:displacement-overview}), such that the user can observe the flow front progression on demand.
  }
\end{figure}

\begin{table}[ht!]
    \caption{Information for the datasets with varying geometries.
        Performance was measured from loading the image files to rendering.}%
    \label{tab:performance-geom}%
    \scriptsize%
    \centering%
    \vspace{0.5em}%
    \begin{tabular}{lccr}%
        \textbf{dataset} & \textbf{resolution} & \textbf{\#frames} & \textbf{time [s]} \\
        \toprule
        circular   & 2448 x 2048 & 232 & 14.24 \\
        octagonal  & 2448 x 2050 & 915 & 50.84 \\
        triangular & 2448 x 2050 &  89 &  9.77
    \end{tabular}
\end{table}

\begin{table}[ht!]
    \caption{Information for the ensemble datasets.
        Performance was measured from loading the image files to rendering.
        Please find the cell legend in the upper right cell of the table.}%
    \label{tab:performance-ensemble}%
    \scriptsize%
    \centering%
    \vspace{0.5em}%
    \newcommand{\tabCa}[1]{\textbf{$\mathbf{\text{Ca}=10^{#1}}$}}%
    \newcommand{\tabM}[1]{\textbf{$\mathbf{\text{M}=#1}$}}%
    \newcommand{\cell}[3]{\begin{minipage}{1.4cm}{\centering#1\\#2\\#3\\}\end{minipage}}%
    \begin{tabular}{l|cccc}%
         & \tabCa{-5} & \tabCa{-4} & \tabCa{-3} & \tabCa{-2} \\
        \toprule
        \tabM{0.2} & \cell{2448 x 2050}{1037}{29.75 s} & \cell{2448 x 2050}{264}{10.58 s} & \cell{2448 x 2050}{618}{18.80 s} & \fbox{\color{gray}\cell{resolution}{\#frames}{time}} \\
        \midrule
        \tabM{1}   & \cell{2448 x 2050}{831}{27.17 s} & \cell{2448 x 2050}{626}{21.76 s} & \cell{2448 x 2050}{232}{ 9.74 s} & \cell{2448 x 2050}{65}{8.53 s} \\
        \midrule
        \tabM{10}  & \cell{2448 x 2050}{2008}{61.88 s} & \cell{2448 x 2050}{1361}{38.67 s} & \cell{2448 x 2050}{320}{11.35 s} & \cell{2448 x 2050}{294}{9.96 s}
    \end{tabular}%
    %
\end{table}

\end{document}